\documentclass{amsart}
\usepackage{graphicx}
\usepackage{natbib,amsfonts,amsmath,verbatim}
\usepackage{amssymb}
\usepackage{rotating}
\usepackage{graphics,psfrag,epsfig,epsf} 
\usepackage{graphicx}
\usepackage[page]{appendix}

\def\baselinestretch{1.5}
  
\def\abstract{\if@twocolumn
\section*{Abstract}
\else \normalsize 
\begin{center}
{\bf Summary\vspace{-.5em}\vspace{0pt}} 
\end{center}
\quotation 
\fi}
\def\endabstract{\if@twocolumn\else\endquotation\fi}

\newcommand{\half}{\mathop{\frac{1}{2}}}

\makeatletter
\@addtoreset{equation}{section}
\makeatother

\newtheorem{remark}{Remark}[section]

\begin{document}
\title[Overdispersed Models]{Variable Selection and Model Averaging in Semiparametric Overdispersed Generalized Linear Models}
\author{Remy Cottet}
\author{Robert Kohn}
\author{David Nott}
\address{University of New South Wales, Sydney, Australia}

\maketitle 

\begin{abstract}
Flexibly modeling the response variance in regression
is important for efficient parameter estimation, correct inference, 
and for understanding the sources of variability in the
response.  Our article considers flexibly modelling
the mean and variance functions within the framework
of double exponential regression models, a class of
overdispersed generalized linear models.  The most
general form of our model describes the mean and dispersion parameters
in terms of additive functions of the predictors.
Each of the additive terms  can be either null, linear, or a fully flexible 
smooth effect.  When the dispersion model is null the mean model is linear 
in the predictors and we obtain a generalized
linear model, whereas with a null dispersion model
and fully flexible smooth terms in the mean model we obtain
a generalized additive model.  Whether or not to include
predictors, whether or not to model their effects linearly
or flexibly, and whether or not to model overdispersion
at all is determined from the data using a fully Bayesian
approach to inference and model selection. Model selection
is accomplished using a hierarchical prior which has many computational
and inferential advantages over priors used in previous empirical
Bayes approaches to similar problems. We describe
an efficient Markov chain Monte Carlo sampling scheme
and priors that make the estimation of the model practical
with a large number of predictors.  The methodology is
illustrated using real and simulated data.  

\noindent  
Key Words: Bayesian analysis; Double exponential family; Hierarchical priors; Variance estimation.
\end{abstract}

\vspace{10pt}

\section{Introduction}
Flexibly modelling the response variance in regression is
important for efficient estimation of mean parameters, correct inference and
for understanding the sources of variability in the response. Response
distributions that are commonly used for modelling non-Gaussian data such as
the binomial and Poisson, although natural and interpretable, have a
variance that is a function of the mean and often real data exhibits more
variability than might be implied by the mean-variance relationship, a
phenomenon referred to as overdispersion. Underdispersion, where the data
exhibits less variability than expected, can also occur, although this is
less frequent. 

Generalized linear models (GLMs) have
traditionally been used to model non-Gaussian regression data (Nelder and
Wedderburn, 1972 \nocite{nelder72}, McCullagh and Nelder, 1989)\nocite
{MCcullagh89}, where the response $y$ has a distribution from the
exponential family and a transformation 
of the mean response is a linear function of predictors. This framework is
extended to generalized additive models (GAMs) by Hastie
and Tibshirani (1990)\nocite{hastie90} where a transformation of the mean is
modelled as a flexible additive function of the predictors. However, the
restriction to the exponential family in GLMs and %
GAMs is sometimes not general enough. While there is
often strong motivation for using exponential family distributions on the
grounds of interpretability, the variance of these distributions is a
function of the mean and the data often exhibit greater variability than is
implied by such mean-variance relationships. 

Quasi-likelihood (Wedderburn, 1974)\nocite{wedderburn74}
provides one simple approach to inference in the presence of overdispersion,
where the exponential family assumption is dropped and only a model for the
mean is given with the response variance a function of the mean up to a
multiplicative constant. However, this approach does not allow
overdispersion to be modelled as a function of covariates. An extension of
quasi-likelihood which allows this is the extended quasi-likelihood of
Nelder and Pregibon (1987)\nocite{nelder87}, but in general extended
quasi-likelihood estimators may not be consistent (Davidian and Carroll,
1988)\nocite{davidian88}. Use of a working normal likelihood for estimating
mean and variance parameters can also be used for modelling overdispersion
(Peck \textit{et al.}, 1984)\nocite{peck84}. However, the non-robustness of
the estimation of mean parameters when the variance function is incorrectly
specified is a difficulty with this approach. Alternatively, a generalized
least squares estimating equation for mean parameters can be combined with a
normal score estimating equation for variance parameters, a procedure
referred to as pseudolikelihood (Davidian and Carroll, 1987)\nocite
{davidian87}. Both the working normal likelihood and pseudolikelihood
approaches are related to the theory of generalized estimating equations
(see Davidian and Giltinan, 1995, p.~57)\nocite{davidian95}. Additive
extensions of generalized estimating equations are considered by Wild and
Yee~(1996)\nocite{wild96}. Smyth (1989)\nocite{smyth89} considers modelling
the mean and variance in a parametric class of models which allows normal,
inverse Gaussian and gamma response distributions, and a quasi-likelihood
extension is also proposed which uses a similar approach to pseudolikelihood
for estimation of variance parameters. Smyth and Verbyla (1999)\nocite
{smyth99} consider extensions of residual maximum likelihood (REML)
estimation of variance parameters to double generalized linear models where
dispersion parameters are modelled linearly in terms of covariates after
transformation by a link function. 

Inference about mean and variance functions using estimating
equations has the drawback that there is no fully specified model, making it
difficult to deal with characteristics of the predictive distribution for a
future response, other than its mean and variance. Model based approaches to
modelling overdispersion include exponential dispersion models and related
approaches (Jorgensen, 1997 \nocite{jorgensen97}, Smyth, 1989), the extended
Poisson process models of Faddy (1997)\nocite{faddy97} and mixture models
such as the beta-binomial, negative binomial and generalized linear mixed
models (Breslow and Clayton, 1993\nocite{breslow93}, Lee and Nelder, 1996) 
\nocite{lee96}. One drawback of mixture models is that they are unable to
model underdispersion. Generalized additive mixed models incorporating
random effects in GAMs are considered by Lin and Zhang
(1999) \nocite{lin99}. Both Yee and Wild (1996) \nocite{Yee96} and Rigby and
Stasinopoulos (2005)\nocite{rigby05} consider very general frameworks for
additive modelling and algorithms for estimating the additive terms. There
is clearly scope for further research on inference: Rigby and Stasinopoulos
(2005) suggest that one use for their methods is as an exploratory tool for
a subsequent fully Bayesian analysis of the kind considered in our article.
See also Brezger and Lang (2005)\nocite{brezger05} and Smith and Kohn (1996)
\nocite{Smith96} for other recent work on Bayesian generalized additive models. 

Our framework for flexibly modelling the mean and variance
functions is based on the double exponential regression models introduced by
Efron (1986), \nocite{efron86} an approach which is also related to the
extended quasi-likelihood of Nelder and Pregibon (1987)\nocite{nelder87}.
The double exponential family has been further extended by Gelfand and
Dalal~(1990) \nocite{Gelfand90a} and Gelfand et al. (1997)\nocite{Gelfand97}%
. Double exponential regression models do not suffer the drawbacks of
extended quasi-likelihood which occur because the extended quasi-likelihood
is not a proper log likelihood function. Semiparametric double exponential
regression models can be used to extend both generalized linear models and
generalized additive models and are able to model both overdispersion and
underdispersion. They provide a convenient framework for modelling as they
have the parsimony and interpretability of GLMs, while
allowing, if necessary, the flexible dependence of the link transformed mean
and variance parameters on predictors. The most general model considered in
our article describes the mean and dispersion parameters after
transformation by link functions as additive functions of the predictors.
For each of the additive terms in the mean and dispersion models we are able
to choose between no effect, a linear effect or a fully flexible effect. For
a null dispersion model and linear effects in the mean model we obtain
generalized linear models, while for a null dispersion model and flexible
effects in the mean model we obtain generalized additive models. The main
contribution of the paper is to describe a fully Bayesian approach to
inference that allows the data to decide whether or not to include
predictors, whether or not to model the effect of predictors flexibly or
linearly, and whether or not to model overdispersion at all. We note that an
important benefit of variable selection and model averaging is that it
produces more efficient model estimates when there are redundant covariates
or parameters. As far as we know, alternative approaches to flexible
modelling of a mean and variance function do not address similar issues of
model selection in a systematic way that is practically feasible when there
are many predictors. Nott~(2004) \nocite{nott04} considers Bayesian
nonparametric estimation of a double exponential family model. However
Nott's paper does not consider model averaging, and
his priors for the unknown functions and smoothing parameters are very
different from those used by our article. 

Our article refines and generalizes the work of Shively et
al.~(1999) \nocite{shively99} and Yau, Kohn and Wood~(2003)\nocite{yau03a}
on nonparametric variable selection and model averaging in probit regression
models. These papers use a data-based prior to carry out variable selection.
To construct this prior it is necessary to first estimate the model with all
flexible terms included, even though the actual fitted model may require
only a much smaller number of such terms. This makes the approach
impractical when there are a moderate to large number of terms in the full
model because the Markov chain simulation breaks down. See Yau et al.
(2003), who discuss this problem and also give some strategies to overcome it.
Another contribution of our article is to overcome the problems with the
data-based prior approach by specifying hierarchical priors for the linear
terms and flexible terms in both the mean and variance models. The
hierarchical specification in our article is also computationally more
efficient than the data-based prior approach because it requires one
simulation run through the data, whereas the data-based approach requires
two runs, the first to obtain the data-based prior and the second to
estimate the parameters of the model. Our approach also applies to variable
selection and model averaging in generalized linear models and overdispersed
generalized linear models where some or all of the predictors enter the
model parametrically. A third contribution of the paper is to develop an
efficient Markov chain Monte Carlo (MCMC) sampling
scheme for carrying out the computations required for inference in the
model. 

The paper is organized as follows. Section 2 describes the
model, priors and our Bayesian approach to inference and model selection.
Section 3 discusses an efficient Markov chain Monte Carlo sampling scheme
for carrying out the calculations required for inference.
Section 4 applies the methodology to both real and simulated data sets.
Section~5 reviews and concludes the paper. 

\section{Model and prior distributions}

\subsection{The double exponential family}

Write the density of a random variable $y$ from a one parameter
exponential family as 
\begin{equation}
{\large p(y;\mu ,\phi /A)=\exp \left( \frac{y\psi -b(\psi )}{\phi /A}+c(y,%
\frac{\phi }{A})\right) \ , }  \label{e:model_expo}
\end{equation}
where $\psi $ is a location parameter, $\phi /A$ is a known scale parameter
and $b(\cdot )$ and $c(\cdot ,\cdot )$ are known functions. The mean of $y$
is $\mu =b^{\prime }(\psi )$ and the variance of $y$ is $(\phi
/A)\,b^{\prime \prime }(\psi )$. This means that $\psi =\psi (\mu )$ is a
function of $\mu $ and so is the variance. Examples of densities which can
be written in this form are the Gaussian, binomial, Poisson and gamma
(McCullagh and Nelder, 1989)\nocite{MCcullagh89}. In \eqref{e:model_expo} we
write the scale parameter as $\phi /A$ in anticipation of later discussion
of regression models where $\phi $ is common to all responses but $A$ may
vary between responses. A double exponential family is defined from a
corresponding one parameter exponential family by 
\begin{equation}
p(y;\mu ,\theta ,\phi /A)=Z(\mu ,\theta ,\phi /A)\theta ^{1/2}p(y;\mu ,\phi
/A)^{\theta }p(y;y,\phi /A)^{1-\theta }\ ,  \label{dexp}
\end{equation}
where $\theta $ is an additional parameter and $Z(\mu ,\theta ,\phi /A)$ is
a normalizing constant. To get some intuition for this definition consider a
Gaussian density with variance 1, and apply the double exponential family
construction: the resulting double Gaussian distribution is in fact an
ordinary Gaussian density with mean $\mu $ and variance $1/\theta $ so that
we can think of the parameter $\theta $ as a scale parameter modelling
overdispersion $(\theta <1)$ or underdispersion ($\theta >1$) with respect
to the original one parameter exponential family density. While the double
Gaussian density is simply the ordinary Gaussian density, for distributions
like the binomial and Poisson, where the variance is a function of the mean,
the corresponding double binomial and double Poisson densities are genuine
extensions which allow modelling of the variance. Efron~(1986) shows that 
\begin{equation}
E(y)\approx \mu ,\,\,\,Var(y)\approx \frac{\phi }{A\theta }b^{\prime \prime
}(\psi ),\quad \mbox{and}\quad Z(\mu ,\theta ,\phi /A)\approx 1\,
\label{e:dexp_approx}
\end{equation}
with these expression being exact for $\theta =1$. Equation (\ref
{e:dexp_approx}) helps to interpret the parameters in the double exponential
model and shows how the GLM mean variance relationship is embedded within
the double exponential family, which is important for parsimonious modelling
of the variance in regression. 
\subsection{ Semiparametric double exponential regression models}

Efron (1986) considers regression models with a response
distribution from a double exponential family, such that both the mean
parameter $\mu $ and the dispersion parameter $\theta $ are functions of the
predictors. Let $y_{1},...,y_{n}$ denote $n$ observed responses, and suppose
that $\mu _{i}$ and $\theta _{i}$ are the location and dispersion parameters
in the distribution of $y_{i}$. For appropriate link functions $g(\cdot )$
and $h(\cdot )$, we consider the model 
\begin{eqnarray}
g(\mu _{i})& = & \ensuremath{\beta^\mu}_{0}+\sum_{j=1}^{p}x_{ij}%
\ensuremath{\beta^\mu}_{j}+\sum_{j=1}^{p}\ensuremath{f^\mu}_{j}(x_{ij})\,
\label{e:dexpreg_g} \\
h(\theta _{i})& = & \ensuremath{\beta^\theta}+\sum_{j=1}^{p}x_{ij}%
\ensuremath{\beta^\theta}_{j}+\sum_{j=1}^{p}f_{j}^{\theta }(x_{ij})\ ,
\label{e:dexpreg_h}
\end{eqnarray}
We first discuss equation \eqref{e:dexpreg_g} for the mean. This equation
has an overall intercept $\ensuremath{\beta^\mu}_{0}$, with the effect of
the $j$th covariate given by the linear term $x_{ij}\ensuremath{\beta^\mu}%
_{j}$ and the nonlinear term $f_{j}^\mu(\ensuremath{x}_{ij})$, which is
modeled flexibly using a cubic smoothing spline prior. Let $%
x_{.,j}=(x_{ij},i=1,\dots ,n)$, for $j=1,\dots ,p$. We standardize each of
the covariate vectors $x_{.,j}$ to have mean 0 and variance 1, which makes
the $x_{.,j},j=1,\dots ,p,$ orthogonal to the intercept term and comparable
in size. This means that the covariates are similar in location and
magnitude, which is important for the hierarchical priors used in our
article. Making the covariates orthogonal to the intercept diminishes the
confounding between the intercept and the covariate. We have also found
empirically that the standardization makes the computation numerically more
stable. 

We now describe the priors on the parameters for the model
given by~\eqref{dexp}, \eqref{e:dexpreg_g} and \eqref{e:dexpreg_h}. The
prior for $\ensuremath{\beta^\mu}_{0}$ is normal but diffuse with zero mean
and variance $10^{10}$. Let $\ensuremath{\beta^\mu}=(\ensuremath{\beta^\mu}%
_{1},...,\ensuremath{\beta^\mu}_{p})^{T}$. To allow the elements of $%
\ensuremath{\beta^\mu}$ to be in or out of the model, we define a vector of
indicator variables $\ensuremath{J^\mu}=(\ensuremath{J^\mu}_{1},\dots ,%
\ensuremath{J^\mu}_{p})$ such that $\ensuremath{J^\mu}_{l}=0$ means that $%
\ensuremath{\beta^\mu}_{l}$ is identically 0, and $\ensuremath{J^\mu}_{l}=1$
otherwise. For given $\ensuremath{J^\mu}$, let $\ensuremath{\beta^\mu}_{J}$
be the subvector of nonzero components of $\ensuremath{\beta^\mu}$, %
i.e. those components $\ensuremath{\beta^\mu}_{l}$
with $\ensuremath{J^\mu}_{l}=1$. We use the notation $N(a,b)$ for the normal
distribution with mean $a$ and variance $b$, $IG(a,b)$ for the inverse gamma
distribution with shape parameter $a$ and scale parameter $b$ and $U(a,b)$
for the uniform distribution on the interval $[a,b]$. With this notation,
the prior on $\ensuremath{\beta^\mu}$, for a given value of $%
\ensuremath{J^\mu}$, is $\ensuremath{\beta^\mu}_{J}|\ensuremath{J^\mu}\sim
N(0,\ensuremath{b^\mu}I)$, where $\ensuremath{b^\mu}\sim IG(s,t)$, $s=101$
and $t=10100.$ This choice of parameters produces an inverse Gamma prior
with mean $101$ and standard deviation $10.15$ which worked well across a range
of examples, both parametric and nonparametric. However, in general the
choice of $s$ and $t$ may depend on the scale and location of the dependent
variable and is left to the user. For a continuous response, standardizing
the dependent variable may be useful here.  The issue of
sensitivity to prior hyperparameters is addressed later 
in the simulations of Section 4.5.  We also assume that $Pr(%
\ensuremath{J^\mu}_{l}=1|\ensuremath{\pi^{\beta\mu}})=\ensuremath{\pi^{\beta%
\mu}}$ for $l=1,...,p$ and that the $J_{l}$ are independent given $%
\ensuremath{\pi^{\beta\mu}}$. The prior for $\ensuremath{\pi^{\beta\mu}}$ is 
$U(0,1)$. 

We now specify the priors for the nonlinear terms $%
\ensuremath{f^\mu}_{j},j=1,\dots ,p$. The discussion below assumes that each 
$x_{.,j}$ is rescaled to the interval $[0,1]$ so that the priors in the
general case are obtained by transforming back to the original scale. Note
that we make the assumption of scaling of the predictors to $[0,1]$ for
expository purposes only to simplify notation in our description of the
priors, since we have previously assumed that our covariates are scaled to
have mean zero and variance one. We assume that the functions ${%
\ensuremath{f^\mu}_{j}}$ are a priori independent of each other, and for any 
$m$ abcissae $z_{1},\dots ,z_{m}$, the vector $(\ensuremath{f^\mu}%
_{j}(z_{1}),\dots ,\ensuremath{f^\mu}_{j}(z_{m}))^{T}$ is normal with zero
mean and with 
\begin{equation*}
\ensuremath{\mbox{cov}}(\ensuremath{f^\mu}_{j}(z),\ensuremath{f^\mu}%
_{j}(z^{\prime }))=\exp (\ensuremath{c^\mu}_{j})\Omega (z,z^{\prime }),
\end{equation*}
where 
\begin{equation}
\Omega (z,z^{\prime })=\half z^{2}(z^{\prime }-\ensuremath{\frac{1}{3}}%
z),\quad \mbox{{for}}\quad 0\leq z\leq z^{\prime }\leq 1.  \label{eq:kernel}
\end{equation}
and $\Omega(z^{\prime },z)=\Omega(z,z^{\prime })$. This prior on $f$ leads
to a cubic smoothing spline for the posterior mean of $\ensuremath{f^\mu}%
_{j} $ (Wahba, 1990, p.~16) with $\exp (\ensuremath{c^\mu}_{j})$ the
smoothing parameter. 

For $j=1,\dots ,p$, let $\ensuremath{f^\mu}_{j}(x_{.,j})=(%
\ensuremath{f^\mu}_{j}(x_{1,j}),\dots ,\ensuremath{f^\mu}_{j}(x_{n,j}))^{T}$%
, and define the $p\times p$ matrix $\ensuremath{V^\mu}_{j}$ as having $%
(i,k) $th element $\Omega (x_{ij},x_{kj})$, so that $\ensuremath{\mbox{cov}}(%
\ensuremath{f^\mu}_{j}(x_{.,j}))=\exp (\ensuremath{c^\mu}_{j})%
\ensuremath{V^\mu}_{j}$. The matrix $\ensuremath{V^\mu}_{j}$ is positive
definite and can be factored as $\ensuremath{V^\mu}_{j}=\ensuremath{Q^\mu}%
_{j}\ensuremath{D^\mu}_{j}{\ensuremath{Q^\mu}_{j}}^{T}$, where $%
\ensuremath{Q^\mu}_{j}$ is an orthogonal matrix of eigenvectors and $%
\ensuremath{D^\mu}_{j}$ is a diagonal matrix of eigenvalues. Let $%
\ensuremath{W^\mu}_{j}=\ensuremath{Q^\mu}_{j}(\ensuremath{D^\mu}_{j})^{\half
}$. Then $\ensuremath{f^\mu}_{j}(x_{.,j})=\ensuremath{W^\mu}_{j}%
\ensuremath{\alpha^\mu}_{j}$, where $\ensuremath{\alpha^\mu}_{j}\sim
N(0,\exp (\ensuremath{c^\mu}_{j})I)$. 

To allow the term $\ensuremath{f^\mu}_{j}$ to be in or out of
the model we introduce the indicator variable $\ensuremath{K^\mu}_{j}$ so
that $\ensuremath{K^\mu}_{j}=0$ means that $\ensuremath{\alpha^\mu}_{j}=0$,
which is equivalent to $\ensuremath{f^\mu}_{j}=0$. Otherwise $%
\ensuremath{K^\mu}_{j}=1$. We also force $\ensuremath{f^\mu}_{j}$ to be null
if the corresponding linear term $\ensuremath{\beta^\mu}_{j}=0$, %
\ensuremath{\mbox{i.e.}}{} if the linear term is zero then we force the
flexible term to also be zero. If $\ensuremath{J^\mu}_{j}=1$, then we assume
that $\ensuremath{K^\mu}_{j}$ is 1 with a probability $\ensuremath{\pi^{f^{%
\mu}}}$, with the prior on $\ensuremath{\pi^{f^{\mu}}}$ uniform. When $%
\ensuremath{K^\mu}_{j}=1$, the prior for $\ensuremath{c^\mu}_{j}\ $is $N(%
\ensuremath{a^{c\mu}},\ensuremath{b^{c\mu}})$, where $\ensuremath{a^{c\mu}}%
\sim N(0,100)$ and $\ensuremath{b^{c\mu}}\sim IG(s,t),$ where $s$ and $t$
are defined above. 

As a practical matter, we order the eigenvalues $%
\ensuremath{D^\mu}_{j}$ of $\ensuremath{V^\mu}_{j}$ in decreasing order and
set to zero all but the largest $m$ eigenvalues, where $m$ is chosen to be
the smallest number such that $\sum_{j=1}^{m}D_{j}^{\mu }/$ $%
\sum_{j=1}^{n}D_{j}^{\mu }\geq 0.98.$ In our work $m$ is usually quite
small, around $3$ or $4$. By setting $D_{j}^{\mu }$ to zero for $j>m,$ we
set the corresponding elements of $\ensuremath{\alpha^\mu}_{j}$ to zero and
it is therefore only necessary to work with an $\ensuremath{\alpha^\mu}_{j}$
that is low dimensional. This achieves a parsimonious parametrization of $%
\ensuremath{f^\mu}_{j}$ while retaining its flexibility as a prior. Our
approach is similar to the pseudospline approach of 
Hastie (1996)\nocite{hastie96} and has
the advantage over other reduced spline basis approaches such as those used
in Eilers and Marx~(1996)\nocite{eilers96}, Yau, Kohn and Wood~(2003) and
Ruppert, Wand and Carroll (2003)\nocite{Ruppert03} of not requiring the
choice of the number or location of knots. 

The interpretation of equation~\eqref{e:dexpreg_h} for the
variance is similar to that of the mean equation. Let $\ensuremath{\beta^%
\theta}=(\ensuremath{\beta^\theta}_{0},\dots ,\ensuremath{\beta^\theta}%
_{p})^{T}$ and define the indicator variable $\ensuremath{J^\theta}=0$ if $%
\ensuremath{\beta^\theta}$ is identically zero, with $\ensuremath{J^\theta}%
=1 $ otherwise, \ensuremath{\mbox{i.e.}}{}, in the variance equation (unlike
the mean equation) all the linear terms are either in or out of the model
simultaneously so we assume that there is linear over or underdispersion in
all the variables or none of them. It would not be difficult to do selection
on the linear terms for individual predictors in the variance model, but we
feel that in many applications it may be possible that there is no
overdispersion, so that the null model where all predictors are excluded
from the variance model is inherently interesting, with inclusion of all
linear terms with a shrinkage prior on coefficients a reasonable
alternative. Our prior parametrizes this comparison directly. When $%
\ensuremath{J^\theta}=1$, we take the prior $\ensuremath{\beta^\theta}\sim
N(0,\ensuremath{b^\theta}I)$, with $\ensuremath{b^\theta}\sim IG(s,t)$ where 
$s$ and $t$ are defined above, and $Pr(\ensuremath{J^\theta}=1)=0.5$.

The hierarchical prior for the nonlinear terms $%
\ensuremath{f^\theta}_{j}$ is similar to that for $\ensuremath{f^\mu}_{j}$.
We write $\ensuremath{f^\theta}_{j}(x_{.,j})=\ensuremath{W^\theta}_{j}%
\ensuremath{\alpha^\theta}_{j}$, with $\ensuremath{\alpha^\theta}_{j}\sim
N(0,\exp (\ensuremath{c^\theta}_{j})I)$. The prior for $\ensuremath{c^\theta}%
_{j}$ is $N(\ensuremath{a^{c\theta}},\ensuremath{b^{c\theta}})$, with $%
\ensuremath{a^{c\theta}}\sim N(0,100),$ $\ensuremath{b^{c\theta}}\sim
IG(s,t) $ where $s$ and $t$ are defined above and $\ensuremath{K^\theta}_{j}$
is 1 with a probability $\ensuremath{\pi^{f^{\theta}}}$, with the prior on $%
\ensuremath{\pi^{f^{\theta}}}$ uniform. We allow the nonlinear terms to be
identically zero by introducing the indicator variables $\ensuremath{K^%
\theta}_{j},j=1,\dots ,p$, where $\ensuremath{K^\theta}_{j}=1$ means that $%
\ensuremath{f^\theta}_{j}$ is in the model and $\ensuremath{K^\theta}_{j}=0$
means that it is not. Similarly to the linear case, we impose that $%
\ensuremath{K^\theta}_{j}=0$ for all $j$ if $\ensuremath{J^\theta}=0$%
\ensuremath{\mbox{, i.e.}}{} if $\ensuremath{J^\theta}=0$ then all the
nonlinear terms in the variance are 0.

This completes the prior specification. The hierarchical prior
is specified in terms of indicator variables that allow selection of linear
or flexible effects for variables in the mean and dispersion models. We
usually use a log link in the dispersion model, $h(\theta )=\log \theta $,
and note that in this case $J^{\theta }=0$ implies that all $\theta $ values
are fixed at one, corresponding to no overdispersion. In some of the
examples below we will sometimes fix $J^{\theta }=0$ which means that our
prior gives a strategy for generalized additive modelling with variable
selection and the ability to choose between linear and flexible effects for
additive terms. 

We note that our framework gives an approach to variable
selection and model averaging in GLM's and overdispersed GLM's by fixing $%
K_{j}^{\mu }=K_{j}^{\theta }=0,j=1,\ldots ,p,$ so that all the terms enter
the model parametrically. The first example of Section 4 illustrates the
ability of our framework to handle situations where a simple parametric
model is appropriate. 

\section{Sampling scheme}

Let $\Delta $ be the set of unknown parameters and latent
variables in the model. We use Markov chain Monte Carlo to obtain the
posterior distributions of functionals of $\Delta $ because in general it is
impossible to obtain these distributions analytically. For an introduction
to Markov chain Monte Carlo methods see e.g. Liu~(2001). The idea of 
\ensuremath{\mbox{Markov
chain Monte Carlo}}{} is to construct a Markov chain $\{\Delta ^{(m)};m\geq
0\}$ such that the posterior distribution is the stationary distribution of
the chain. By running the chain a long time from an arbitrary starting value 
$\Delta ^{(0)},$ and after discarding an initial ``burn in'' sequence of $b$
iterations say, where the distribution of the state is influenced by $\Delta
^{(0)}$, we will be able to obtain approximate dependent samples from $%
p(\Delta |y)$. One estimator of a posterior expectation of interest, $%
E(h(\Delta )|y)$, is 
\begin{equation*}
\frac{1}{s}\sum_{i=b+1}^{b+s}h(\Delta ^{(i)}),
\end{equation*}
where the first $b$ iterations are discarded in taking the sample path
average. 

When generating the elements of $\ensuremath{J^\mu}$ it is
useful to analytically integrate out $\ensuremath{\pi^{\beta\mu}}$ in the
prior for $\ensuremath{J^\mu}$ to obtain 
\begin{align*}
p(\ensuremath{J^\mu}) & = B(1+\sum_l \ensuremath{J^\mu}_l,1+\sum_l (1-%
\ensuremath{J^\mu}_l))
\end{align*}
where $B(\cdot,\cdot)$ is the Beta function. Similar remarks apply to $%
\ensuremath{K^\mu}$ and $\ensuremath{\pi^{f\mu}}$ and $\ensuremath{K^\theta}$
and $\ensuremath{\pi^{f\theta}}$. Thus $\ensuremath{\pi^{\beta\mu}}, %
\ensuremath{\pi^{f\mu}}$ and $\ensuremath{\pi^{f\theta}}$ do not appear in
the sampling scheme below. 

The sampling scheme cycles between different kernels for
updating subsets of the parameters to construct the transition kernel for
the Markov chain. The kernels for updating the subsets are standard Gibbs
and Metropolis-Hastings kernels (see Liu, 2001, for further background) 
\nocite{liu01}. An update of $\Delta $ at a given iteration of our sampling
scheme proceeds in the following steps, with further details given in the
Appendix: 

\begin{enumerate}
\item  {\normalsize Sample $\ensuremath{\beta^\mu}_{0}$. }

\item  {\normalsize For $j=1,...,p$, sample $(\ensuremath{\beta^\mu}_{j},%
\ensuremath{J^\mu}_{j})$ as a block. }

\item  {\normalsize Sample $\ensuremath{b^\mu}$. }

\item  {\normalsize For $j=1,...,p$, sample $(\ensuremath{\alpha^\mu}_{j},%
\ensuremath{c^\mu}_{j},\ensuremath{K^\mu}_{j})$ as a block. }

\item  {\normalsize Sample $\ensuremath{a^{c\mu}}$, $\ensuremath{b^{c\mu}}$. 
}

\item  {\normalsize Sample $(\ensuremath{\beta^\theta},\ensuremath{J^\theta})
$ as a block. }

\item  {\normalsize Sample $\ensuremath{b^\theta}$. }

\item  {\normalsize For $j=1,...,q$, sample $(\ensuremath{\alpha^\theta}_{j},%
\ensuremath{c^\theta}_{j},\ensuremath{K^\theta}_{j})$ as a block. }

\item  {\normalsize Sample $\ensuremath{a^{c\theta}},\ensuremath{b^{c\theta}}%
.$ }
\end{enumerate}

At each step the update of a block of parameters is done
conditional on the current values for the remaining parameters. 

We briefly discuss one important computational issue arising in
implementation of the sampling steps above, namely the calculation of the
normalizing constants $Z(\mu ,\theta ,\phi /A)$ in (\ref{dexp}).
Calculation of this quantity is needed in order to
calculate the log-likelihood. 
Our approach is simply to calculate $Z(\mu ,\theta ,\phi /A)$ on a fine grid
for $\mu $ and $\theta $ and to then use an interpolation scheme for values
of $\mu $ and $\theta $ not on the grid. In the case of the double Poisson
model we sum the density up to a large truncation point where the
contribution of remaining terms to the sum is negligible in order to
calculate the normalizing constant at the grid points.
This is done offline before beginning the MCMC calculations. For the
examples discussed later we used a grid for $g(\mu _{i})$ and $h(\theta _{i})
$ extending from log(-50) to log(50) in steps of 2. The summation of the
density was truncated at 1000 in calculation of the normalizing constants
for the double poisson case. For the double binomial case, we considered the
same grid for $g(\mu _{i})$ and $h(\theta _{i})$. Here we calculate the
normalizing constant on this grid for all possible values of the weight $%
\phi /A$ -- the possible weight values are $1/n_{i}$, $i=1,...,n$ where $n_{i}
$ is the number of trials for the $i$th observation. Efron (1986) also
describes some asymptotic approximations for the normalizing constant 
but we do not use these approximations here.

\section{Empirical Results}

This section illustrates the application of our methodology in
a range of examples, starting in Section 4.1 with a simple parametric
example involving overdispersed count data. Section 4.2 considers flexible
GAM modelling with variable selection for binary data. Section 4.3 considers
the same data as in Section 4.2, but incorporates flexible modelling of
interaction terms and variable selection on the interactions. This example
illustrates that our methodology can handle very large problems where
previously proposed empirical Bayes approaches are infeasible. Section 4.4
considers flexible modelling of overdispersed count data, and Section 4.5
summarizes a simulation study examining the frequentist performance of our
method.

\subsection{Fully parametric regression}

This section illustrates our variable selection methodology in
a parametric setting by fitting an overdispersed Poisson model to the pox
lesions chick data (http://www.statsci.org/-data/general/pocklesi.html). We
think that it is important to start with consideration of a simple fully
parametric example to emphasize that our methodology can be applied with the
flexible terms excluded for small data sets where it may only be feasible to
fit a simple parametric model. Such small data sets are reasonably common in
applications of overdispersed generalized linear models.

\begin{figure}[h]
\begin{center}
{\normalsize \epsfig{file=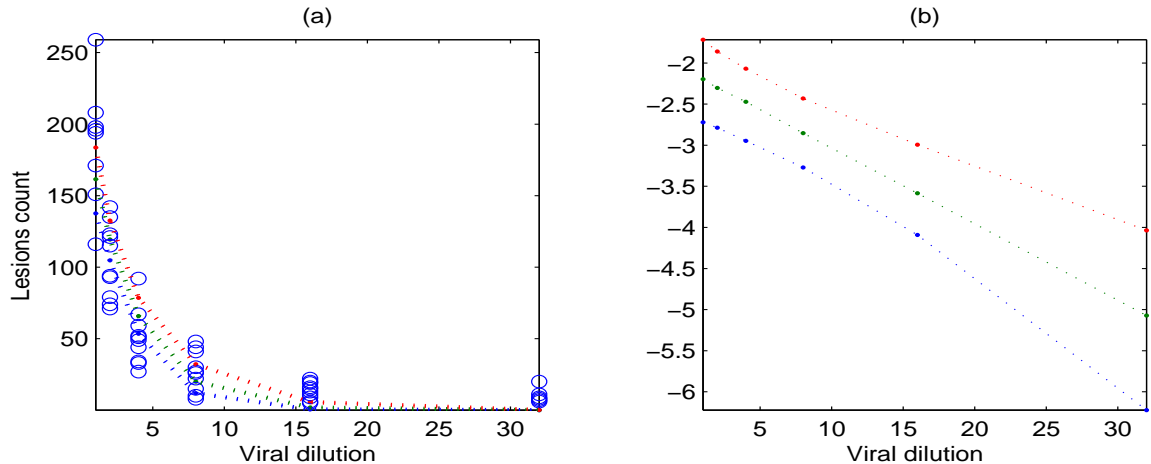,width={\textwidth},height=2.5in}  }
\end{center}
\caption{Lesions on chick embryos. Plot of the estimated posterior means of
mean and variance parameters as a function of viral dilution together with
pointwise 95$\%$ credible intervals. Panel (a) plots estimated values of the
parameter $\protect\mu $ along with the data and panel (b) plots estimated
values of $\log \protect\theta $.}
\label{fig:chick}
\end{figure}

The dependent variable is the counts of lesions produced on
membranes of chick embryos by viruses of the pox group while the independent
variable is the level of dilution of the viral medium. There are $58$
observations in this data set. The data is analysed in Breslow (1990)\nocite
{breslow90} and Podlich, Faddy and Smyth (2004)\nocite{Podlich04}. In our
model $g(\mu _{i})=\log (\mu _{i})$ and $h(\theta _{i})=\log (\theta _{i}),$
with $g$ and $h$ linear functions of the dilution level. Figure \ref
{fig:chick} plots the fit for the parameter $\mu $ and for $\log \theta $ as
a function of viral dilution. The posterior probability of overdispersion
(that is, the posterior probability of $J_\theta=1$) is approximately one
strongly suggesting that there is overdispersion, and this is consistent
with previous analyses by Breslow (1990) and Podlich, Faddy and Smyth
(2004). The results show that overdispersion is increasing with an increase
of the viral dilution while the lesions count decreases with dilution. 

\begin{figure}[ht!]
\begin{center}
{\normalsize 
\epsfig{file=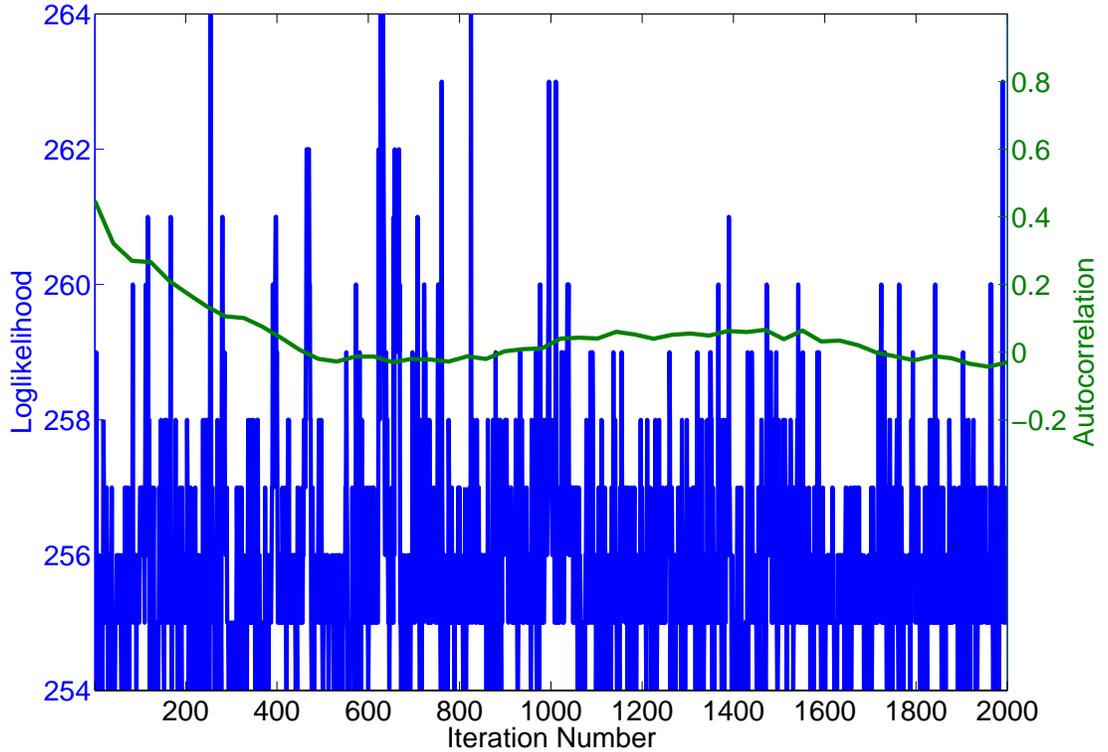,width={\textwidth},width=4in,angle=-90}
}
\end{center}
\caption{Lesions on chick embryos. Plot of the log-likelihood versus
iteration and estimated autocorrelation based on 2000 iterations with
2000 burn in.}
\label{fig:chickdiag}
\end{figure}

Figure \ref{fig:chickdiag} shows a plot of the log-likelihood
versus iteration number in our MCMC sampling scheme as well as the
autocorrelation function of the log-likelihood values based on 2000
iterations with 2000 burn in. These plots show that our sampling scheme
converges rapidly and mixes well. Corresponding plots for our other examples
(not shown) confirm the excellent properties of our sampling scheme. The
4000 iterations of our sampler took 280 seconds on a machine with 2.8 GHz
processor.  For all the examples considered in this paper 
programs implementing our sampler were written in Fortran 90.

\subsection{ Binary logistic regression}

This section considers the Pima Indian diabetes dataset
obtained from the UCI repository of machine learning databases
(http://www.ics.uci.ecu/MLRepository.html). The data is analysed by Wahba,
Gu, Wang and Chapell (1995)\nocite{wahba95}. A population of women who were
at least 21 years old, of Pima Indian heritage and living near Phoenix,
Arizona, was tested for diabetes according to World Health Organization
criteria. The data were collected by the US National Institute of Diabetes
and Digestive and Kidney Diseases. 724 complete records are used after
dropping the aberrant cases (as in Yau et al. 2003). The dependent variable
is diabetic or not according to WHO criteria, where a positive test is coded
as ``1''. There are eight covariates: number of pregnancies, plasma glucose
concentration in an oral glucose tolerance test, diastolic blood pressure
(mm Hg), skin triceps skin fold thickness (mm), 2-Hour serum insulin (mu
U/ml), body mass index (weight in kg/(height in m)$^{2}$), diabetes pedigree
function, and age in years.

\begin{figure}[ht!]
\begin{center}
\epsfig{file=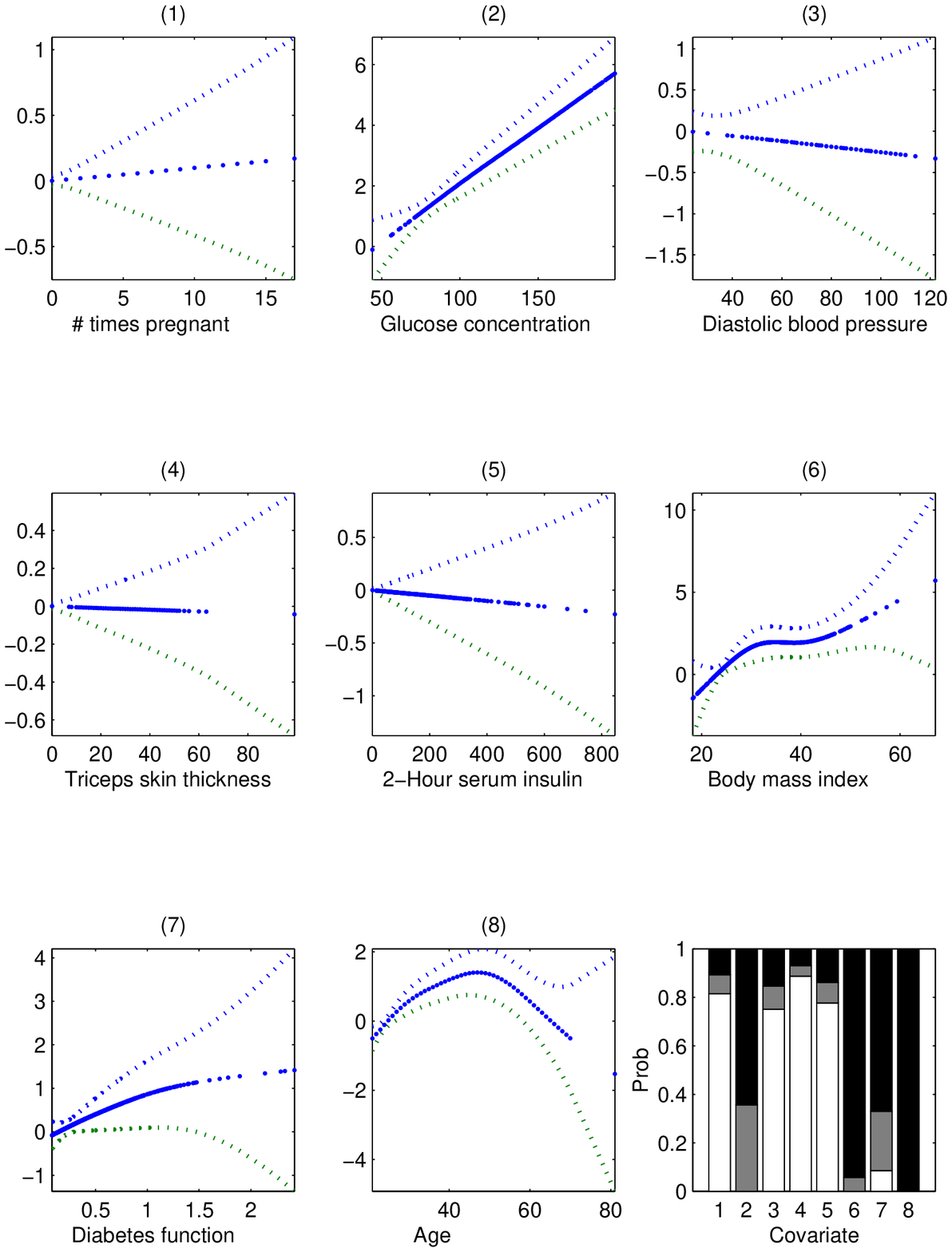,width={\textwidth},height=7in} 
\end{center}
\caption{Logistic Diabetes data. Plots of the posterior means of the
covariate effects at the design points (dotted bold line) and 95\% credible
intervals (dotted lines). The barplot gives the posterior probability of
each covariate function being null (white), linear (grey) and flexible
(black).}
\label{fig:Diabetesmuflex}
\end{figure}

This section fits a main effects binary logistic regression to
the data. In the framework of section 2, we are fixing all $\theta _{i}$ at $%
1$ and so are fitting a generalized linear additive model with $g(\mu
_{i})=\log (\mu _{i}/(1-\mu _{i}))$ that allows for variable selection and
choice between flexible and linear effects for the additive terms. The
results are shown in figure \ref{fig:Diabetesmuflex}, with the barplot
showing the posterior probabilities of effects for each predictor being
null, linear and flexible. The barplot suggests that the number of
pregnancies, diastolic blood pressure, skin triceps skin fold thickness and
2-Hour serum insulin do not seem to help predict the occurrence of diabetes
when the other covariates are in the model. Figure \ref{fig:Diabetesmuflex}
also shows that plasma glucose concentration has a strong positive linear
effect, and body mass index, diabetes pedigree function and age have
nonlinear effects. 

\begin{figure}[ht!]
\begin{center}
\epsfig{file=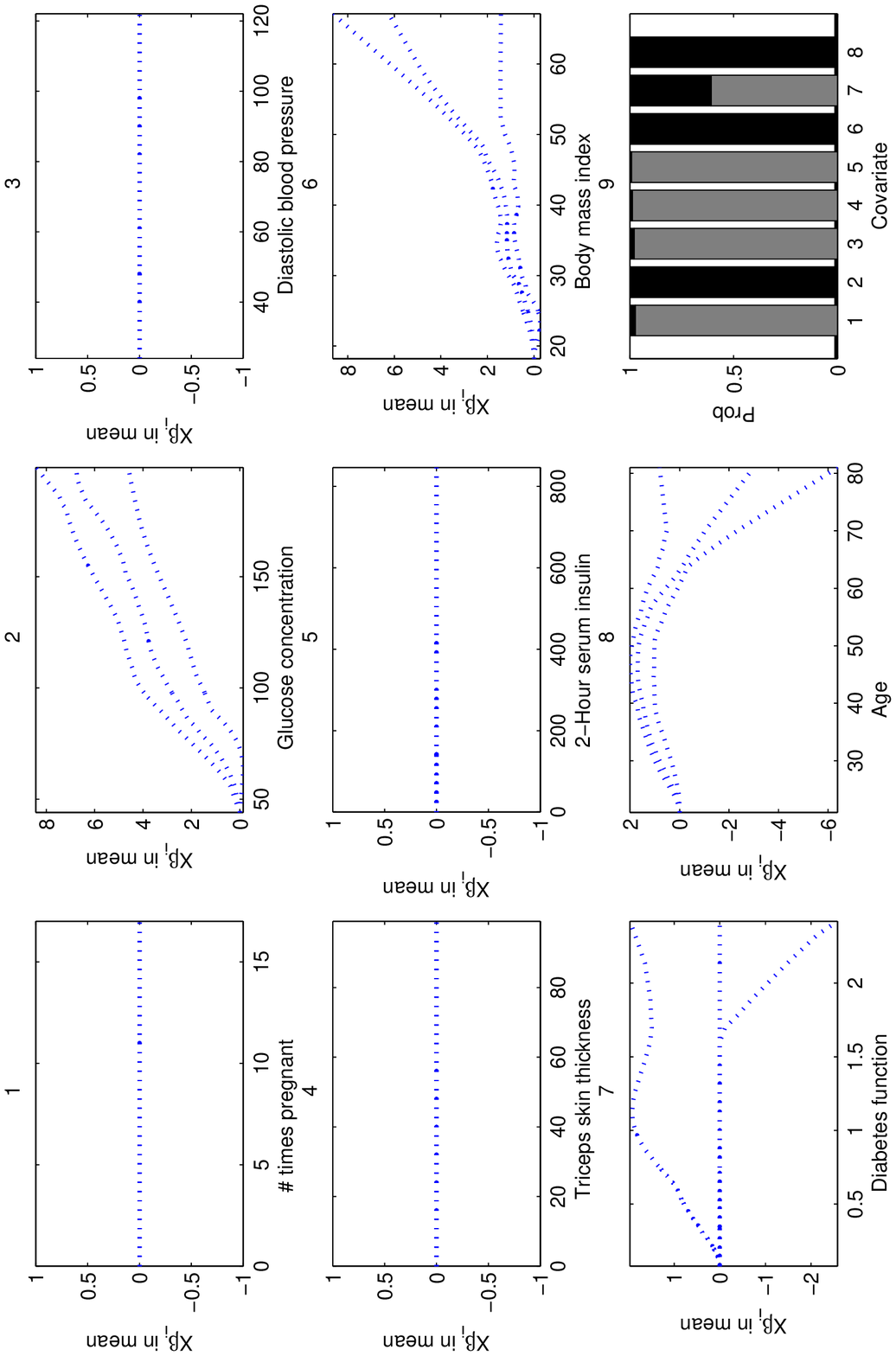,width={\textwidth},height=6in,angle=-90}
\end{center}
\caption{Logistic Diabetes data. Plots of the posterior means of the
covariate effects at the design points (dotted bold line) and 95\% credible
intervals (dotted lines) for data based priors approach. The barplot gives
the posterior probability of each covariate function being in or out of the
model. }
\label{fig:Diabetesmuflexdbp}
\end{figure}

Our method extends the approach of Yau et al. (2003)\nocite
{yau03a} to any GAM whereas Yau et al. (2003) rely on the probit link to
turn a binary regression into a regression with Gaussian errors. Our
approach has several other advantages over Yau et al. (2003) as explained in
the introduction and section 4.3. Figure \ref{fig:Diabetesmuflexdbp} shows
the results of applying a variant of the data-based priors approach of Yau 
\textit{et al.} (2003) to the diabetes data. 

Since Yau \textit{et al.} (2003) did not consider logistic regression we
need to explain how the data-based priors approach was applied here. First,
the full model was fitted (all flexible terms included) with a
noninformative but proper prior on the $c_{j}^{\mu }$ parameters ($IG(s,t)$
with $s=27$ and $t=1300$). Linear terms are selected together with flexible
terms in the approach of Yau \textit{et al.} (2003). The posterior medians
and variances for the $c_{j}^{\mu }$ (with the variances inflated by a
factor of $n$) in this fit of the full model were then used to set means and
variances for a normal prior on the $c_{j}^{\mu }$ in our variable selection
prior similar to Yau \textit{et al.} (2003). The results of Figure \ref
{fig:Diabetesmuflexdbp} are similar to those shown in Figure \ref
{fig:Diabetesmuflex} but as we have already discussed the data based priors
approach is not feasible in general for doing selection with a large number
of terms. Note that in the barplot we only show posterior probabilities for
covariate effects being in or out of the model, since linear and flexible
terms are selected together in the approach of Yau \textit{et al.} (2003).

\begin{figure}[th!]
\begin{center} 
\epsfig{file=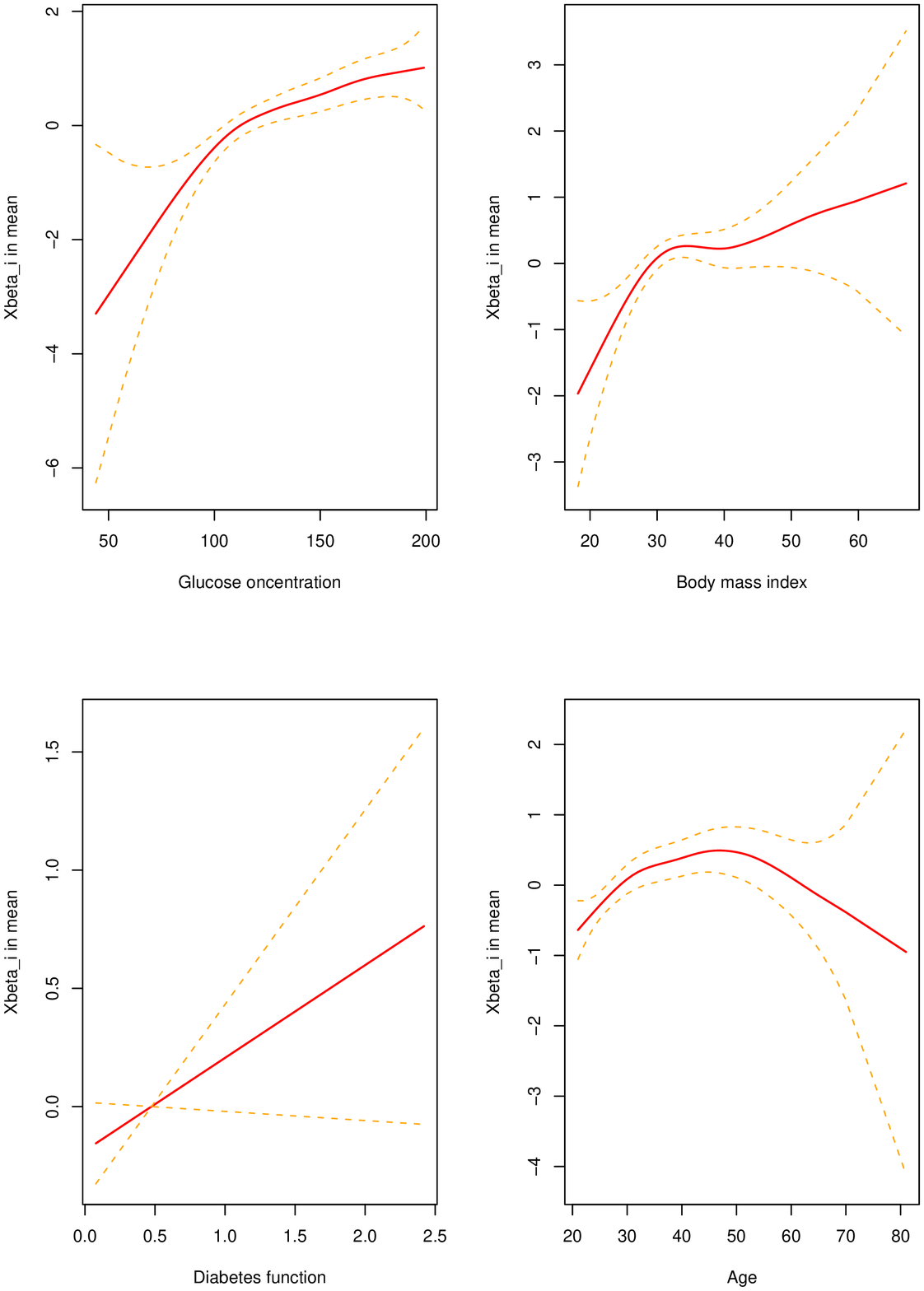,width={\textwidth},height=7in}
\end{center}
\caption{Logistic Diabetes data. Plots of the estimated covariate effects 
(solid line) and 95\% credible intervals (dotted
lines) using GAMLSS and a backward stepwise model selection procedure with
generalized AIC criterion for model selection. }
\label{fig:Diabetesmuflexgamlss}
\end{figure}

We have also compared our approach to that implemented in the
GAMLSS R package of Rigby and Stasinopoulos (2005). We implemented a
backward stepwise model selection procedure to select between null, linear
and flexible effects for the predictors starting with the model containing
all flexible terms. We use their generalized AIC criterion with 
their penalty parameter \# set to $2$
to compare models in the backward stepwise procedure. The final model has
flexible terms for Age, Glucose concentration and Body mass index, and a linear
term for diabetes function. The fit is shown
in Figure \ref{fig:Diabetesmuflexgamlss}. 

We also report for this example the acceptance rates for our
Metropolis-Hastings proposal at step 4 of our sampling scheme in the case
where there is currently a flexible term in the model for a predictor and a
flexible term is also included in the proposed model. We report acceptance
rates for each predictor where there is posterior probability greater than $%
0.1$ of inclusion of a flexible term. The corresponding acceptance rates are
19\%, 11\% and 38\% for the three effects selected. These acceptance rates help
to quantify the usefulness of Approximation 2 decribed in the appendix for
constructing the Metropolis-Hastings proposal. Results for this example were
based on 4000 iterations of our sampling scheme with 4000 burn in. The 4000
iterations took $1300$ seconds on a machine with 2.8 GHz
processor.

\subsection{High dimensional binary logistic regression}

This section extends the model for the Pima Indian dataset to
allow for flexible second order interactions. This means that the model
potentially has 36 flexible terms, 8 main effects and 28 interactions. The
purpose of this section is to show how our class of models can handle
interactions and that the hierarchical priors allow variable selection with
a large number of terms. This is infeasible with the data-based prior
approach of Yau {\it et al.} (2003)\nocite{yau03a}, as explained in section~1. 

We write the generalization of the mean model (2.4) as 
\begin{equation*}
g(\mu _{i})=\beta _{0}+\sum_{j=1}^{p}x_{ij}\beta
_{j}+\sum_{j=1}^{p}f_{j}^{M}(x_{ij})+\sum_{j=1}^{p}%
\sum_{k<j}f_{jk}^{I}(x_{ij},x_{ik})\ .
\end{equation*}
We have dropped the superscript $\mu $ from the $\beta _{j}$ and the $f_{j}$
because we are dealing with the mean equation only. However, we write the
flexible main effects and interactions as $f_{j}^{M}$ and $f_{jk}^{I}$,
where $M$ means main effect and $I$ means an interaction. The prior for the $%
f_{j}^{M}$ is the same as for the flexible main effects in Section~2. For
the interaction effects we assume that any collection of $%
\{f_{jk}^{I}(x_{i},z_{i}),i=1,\dots ,m\}$ is Gaussian with zero mean and 
\begin{equation*}
\mbox{cov}(f_{jk}(x,z),f_{jk}(x^{\prime },z^{\prime }))=\exp
(c_{jk}^{I})\Omega (x,x^{\prime })\Omega (z,z^{\prime })\ ,
\end{equation*}
where $\Omega (z,z^{\prime })$ is defined by equation (\ref{eq:kernel}).
This gives a covariance kernel for the $f_{jk}$ that is the tensor product
of univariate covariance kernels (Gu 2002, section 2.4)\nocite{gu02}. Once
the covariance matrix for $(f_{jk}^{I}(x_{ij},x_{ik}),i=1,\dots ,n)$ is
constructed, we factor it to get a parsimonious representation as in
Section~2. The smoothing parameters $c_{jk}^{I}$ have a similar prior to the 
$c_{j}^{\mu }$ in Section~2. To allow for variable selection of the flexible
main effects, let $K_{j}^{M}$ be indicator variables such that $K_{j}^{M}=0$
if $f_{j}^{M}$ is null and $K_{j}^{M}=1$ otherwise. The prior for $K_{j}^{M}$
is the same as for the $K_{j}^{\mu }$ in Section~2. To allow variable
selection on the flexible interaction terms, let $K_{jk}^{I}$ be an
indicator variable which is 0 if $f_{jk}^{I}$ is null, and is 1 otherwise.
To make the bivariate interactions interpretable, we only allow a flexible
interaction between the $j$th and $k$th variables if both the flexible main
effects are in, i.e., if $K_{j}^{M}=0$ or $K_{k}^{M}=0$ or both, then $%
K_{jk}^{I}=0$. If both $K_{j}^{M}$ and $K_{k}^{M}$ are 1, then 
\begin{equation*}
p(K_{jk}^{I}=1|K_{j}^{M}, K_{k}^{M},\pi ^{I})=\pi ^{I}
\end{equation*}
where $\pi ^{I}$ is uniformly distributed. The generation of the interaction
effects parameters $(\alpha _{jk}^{I},c_{jk}^{I},K_{jk}^{I})$ is similar to
the generation of the other parameters in the model. First the indicator
variable is generated from the prior $p(K_{jk}^{I}|K_{j}^{M}K_{k}^{M})$.  If $%
K_{jk}^{I}=1$ then $\alpha _{jk}^{I},c_{jk}^{I}$ are generated as described
in the appendix for the generation of the other parameters, otherwise $%
\alpha _{jk}^{I}$ is set to zero. 

No interactions were detected when the interaction model were
fitted to the data. To test the effectiveness of the methodology at
detecting interactions we also generated observations from the estimated
main effects model, but added an interaction between Diabetes pedigree
function and Age. Writing $x$ and $z$ respectively for these two predictors
the interaction term added to our fitted additive model for $\log \mu
_{i}/(1-\mu _{i})$ in the simulation takes the simple
multiplicative form $xz$.
Table \ref{tab:postprobflex} reports the results of the estimation when the
interaction model was fitted to the artificial data, and shows that the
interaction effect between variables $7$ and $8$ is detected. 

\begin{table}[h]
\begin{center}
\begin{tabular}{c|cccccccc}
\hline\hline
& \multicolumn{8}{c}{Covariate} \\ 
& 1 & 2 & 3 & 4 & 5 & 6 & 7 & 8 \\ \hline
Null & 0.88 & 0.00 & 0.86 & 0.81 & 0.76 & 0.03 & 0.00 & 0.00 \\ 
Linear & 0.08 & 0.58 & 0.08 & 0.05 & 0.11 & 0.53 & 0.00 & 0.00 \\ \hline
1 & 0.00 & 0.00 & 0.00 & 0.00 & 0.00 & 0.01 & 0.03 & 0.02 \\ 
2 &  & 0.42 & 0.01 & 0.03 & 0.05 & 0.13 & 0.24 & 0.21 \\ 
3 &  &  & 0.06 & 0.02 & 0.02 & 0.00 & 0.02 & 0.03 \\ 
4 &  &  &  & 0.14 & 0.04 & 0.03 & 0.07 & 0.07 \\ 
5 &  &  &  &  & 0.13 & 0.03 & 0.06 & 0.06 \\ 
6 &  &  &  &  &  & 0.44 & 0.023 & 0.23 \\ 
7 &  &  &  &  &  &  & 1.00 & 1.00 \\ 
8 &  &  &  &  &  &  &  & 1.00 \\ \hline\hline
\end{tabular}
\vspace{10pt}
\caption{Simulated diabetes data with interaction. Posterior probabilities
of null, linear and flexible main effects and flexible interaction effects.
The table is interpreted as follows for covariate $4.$ The posterior
probabilities of a null, linear and flexible main effect are $0.81$, $0.05$
and $0.14.$ The posterior probability of a flexible interaction between
covariates $3$ and $4$ is 0.02. Other entries in the table are interpreted
similarly.}
\end{center}
\label{tab:postprobflex}
\end{table}

\subsection{Double Binomial model}

This example considers a dataset in Moore and Tsiatis (1991) 
\nocite{moore91} and analyzed by Aerts and Claeskens (1997)\nocite{aerts97}
using a local beta binomial model. An iron supplement was given to 58 female
rats at various dose levels. The rats were then made pregnant and sacrificed
after 3 weeks. The litter size and the number of fetuses dead were recorded
as well as the hemoglobin levels of the mothers. We fitted a double binomial
model to the data to try to explain the proportion of dead foetuses with the
level of hemoglobin of the mother and litter size as covariates.

\begin{figure}[ht!]
\centering
\epsfig{file=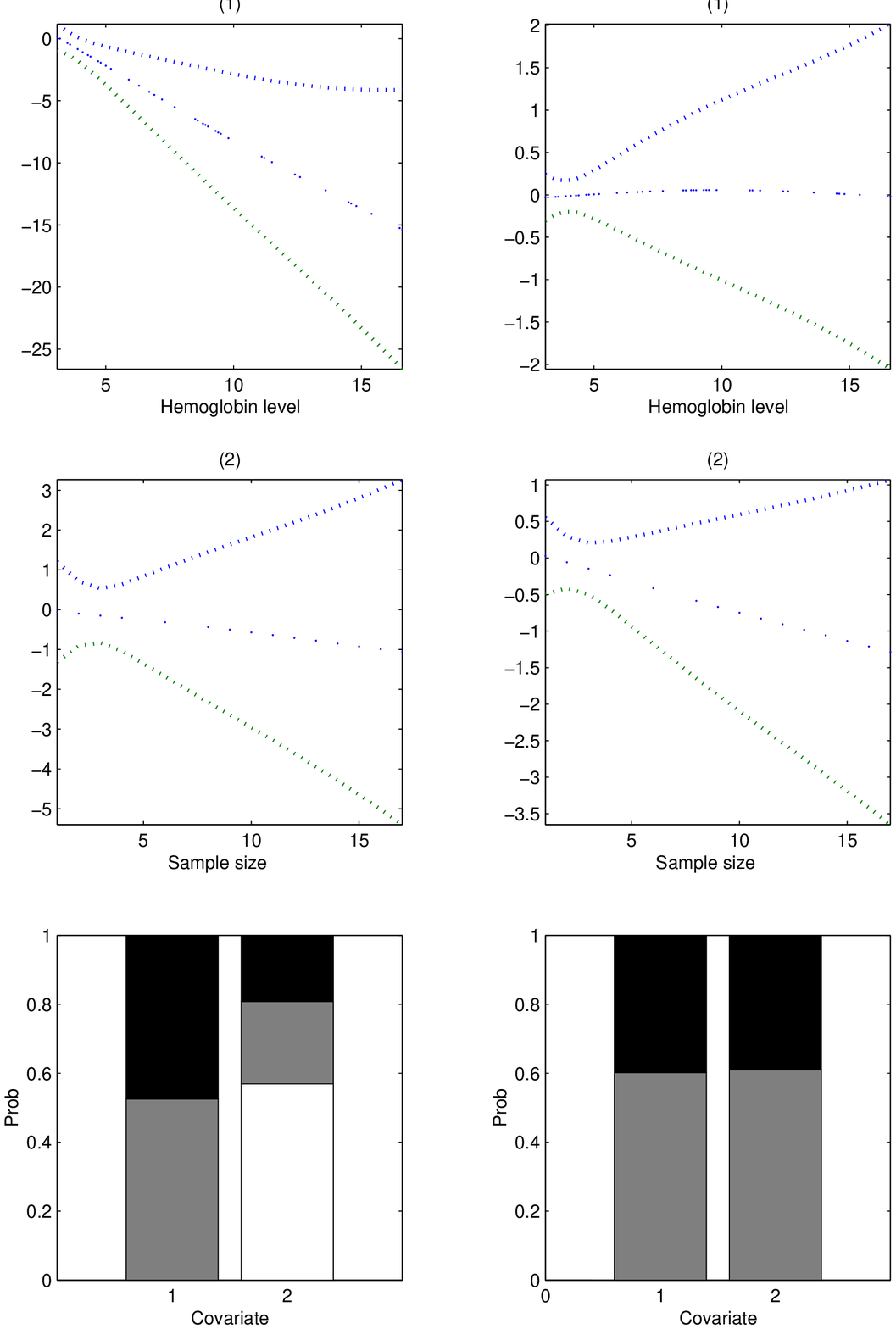,width={\textwidth},height=6in} 
\caption{Double binomial rats data. Left column: Plot of the posterior means
of the effects in the mean model (bold dotted line) 
together with the 95$%
\%$ credible intervals (dotted lines). The barplot plots the posterior
probability of the effects being null (white), linear (grey) and flexible
(black). Effects for the dispersion component are plotted in the right
column. }
\label{fig:ratsmuflex}
\end{figure}

Figures \ref{fig:ratsmuflex} summarizes the estimation results
and shows the presence of overdispersion. As usual when dealing with
binomial like data the count response is rescaled to be a proportion, so
that the parameter $\mu $ here is on the scale $[0,1]$. We have used a
logistic link for $\mu $ and a log link for $\theta $. The results suggest
there is no effect for sample size in the mean model, with some support for
either linear or flexible effects for hemoglobin in the mean and variance
models and for sample size in the variance model.

Similar to the diabetes example, we report the acceptance rates
for our Metropolis-Hastings proposals at steps 4 and 8 of our sampling
scheme in the case where there is currently a flexible term in the model for
a predictor and a flexible term is also included in the proposed model. We
report acceptance rates for each predictor and both the mean and the
variance model where there is posterior probability greater than $0.1$ of
inclusion of a flexible term. The acceptance rates for the
mean model are 2.5\% for hemoglobin and 2.76\% for litter size.
For the variance model, no flexible effect was selected. 
Although the acceptance rates are quite low here, our proposals are
still good enough to obtain reasonable mixing. Results for this example were
based on 5000 iterations of our sampling scheme with 5000 burn in. The 5000
iterations took 4039 seconds on a machine with 2.8 GHz processor.

For this example we also compare an implementation of our methodology
using a beta-binomial response distribution to flexible beta-binomial
regression implemented in the GAMLSS library in R (Rigby and 
Stasinopoulos, 2005).  Implementation of our method for the
beta-binomial family rather than the double exponential is straightforward
as our computational scheme makes no particular use of the double
exponential family assumption, but only the idea of mean and variance
parameters being modelled flexibly as a function of covariates.  
For beta-binomial regression, Rigby and Stasinopoulos (2005)
parametrize the model in terms of a mean parameter $\mu$ and dispersion
parameter $\sigma$ which is $\rho/(1-\rho)$ where $\rho$ is 
the intracluster correlation 
(if we regard each count observation as an
observation of a sequence of exchangeable binary random variables, 
the intracluster correlation is just the correlation between
a pair of these binary random variables).  
Large value of $\sigma$ correspond
to overdispersion, whereas $\sigma=0$ corresponds to no
overdispersion.   Our model is similar to before, excpet that we replace
our model for $h(\theta_i)$ in (\ref{e:dexpreg_h}) with a model of
the same form for $h(\sigma_i)$ where $\sigma_i$ is the dispersion
parameter for observation $i$ and $h(\cdot)$ is a link function
which we choose here as the log function.  

\begin{figure}[ht!]
\begin{center}
{\normalsize \epsfig{file=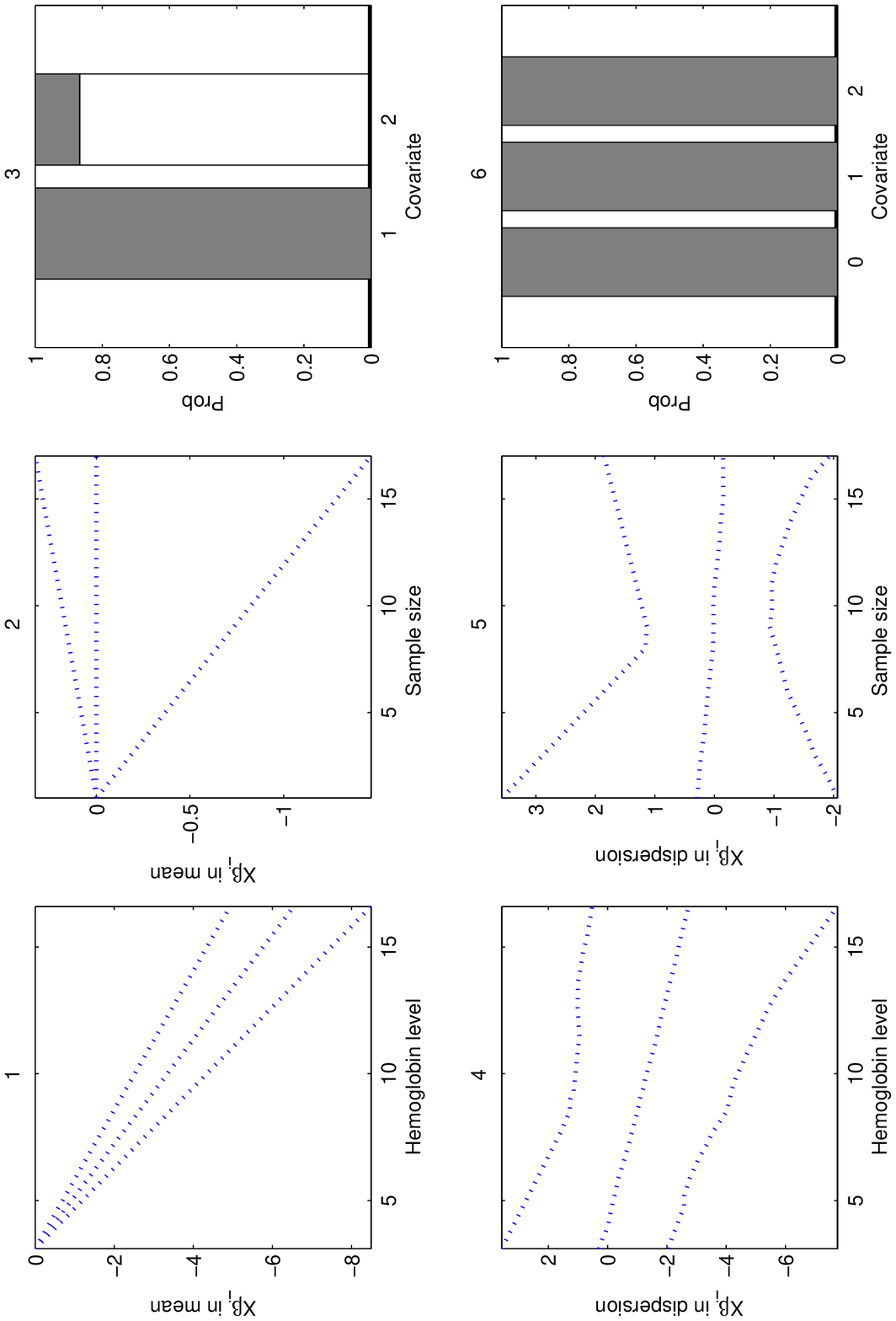,width={\textwidth},height=6in,angle=-90}  }
\end{center}
\caption{
Plots of effects for hemoglobin
in the mean and dispersion models (left top and bottom respectively)
and for sample size (middle top
and bottom respectively)
together
with pointwise 95\% credible intervals (dotted lines) for rats data.  
On the right, the barplots show the probabilities for null (white), 
linear (grey) and 
flexible (black) effects.   
}
\label{fig:betabinratsmuflex}
\end{figure}

\begin{figure}[ht!]
\begin{center}
\epsfig{file=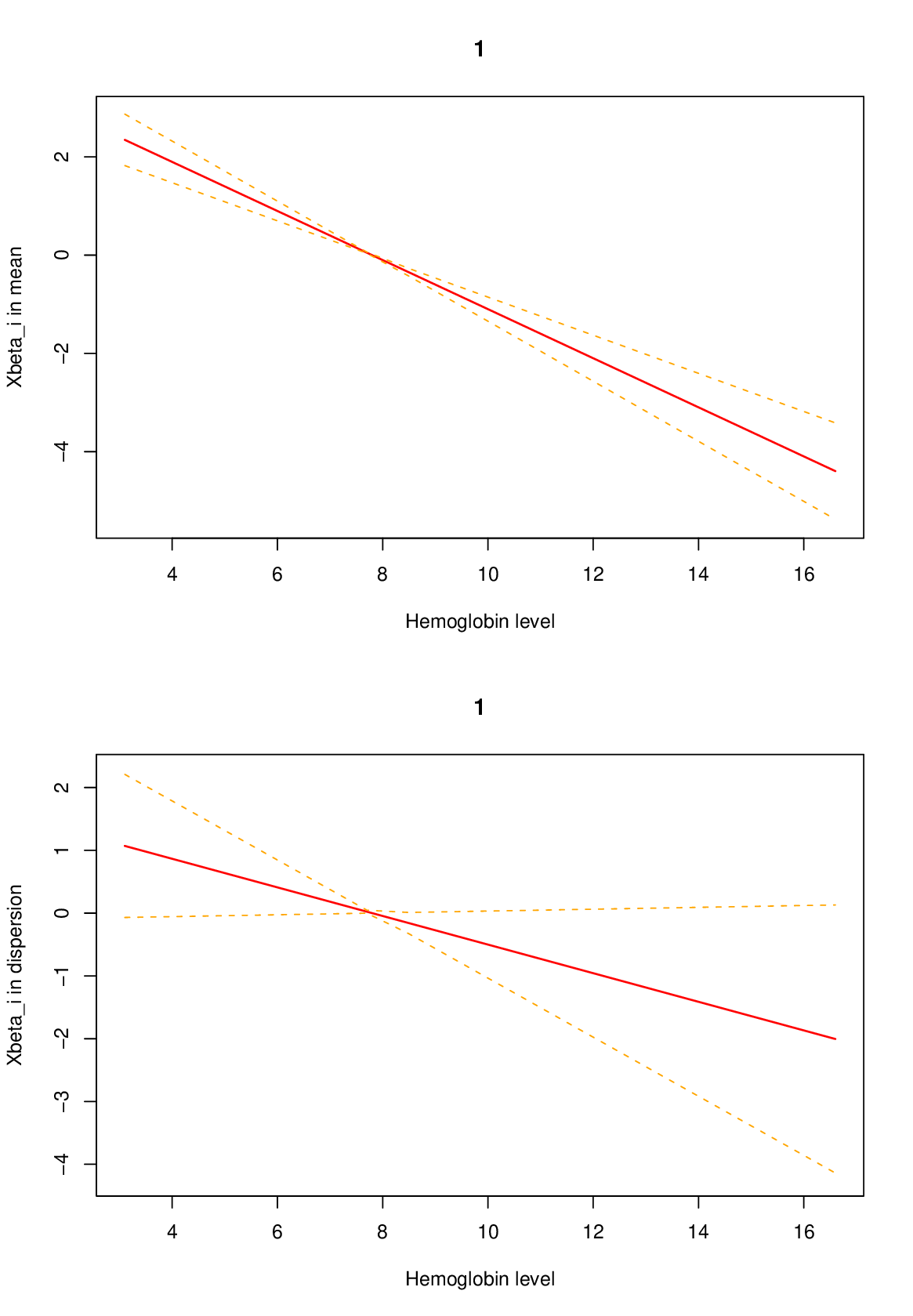,width={\textwidth},height=6in}  
\end{center}
\caption{
Plots of effects for hemoglobin
in the mean model and dispersion models (top and bottom 
respecitvely)
together with pointwise 95\% credible intervals (dotted lines) for
rats data where the fit is obtained from the GAMLSS package.  
}
\label{fig:betabingamlssratsmuflex}
\end{figure}

Figures \ref{fig:betabinratsmuflex} and 
\ref{fig:betabingamlssratsmuflex} show the results of our fit and the GAMLSS fit
(with all terms flexible) for the rat data.  We can see that
the fits are similar.

One advantage of our approach is greater 
computational stability, 
a feature that we believe is related to our shrinkage priors.  
We simulated several
datasets from our fitted model for the mean, but assuming no overdispersion
($\sigma=0$) and then attempted to fit to this simulated data using
GAMLSS and our Bayesian approach with a beta-binomial model.  
The Bayesian approach produces
satisfactory results, but attempting to fit the model in GAMLSS
even with only an intercept and no covariates in the variance
model results in convergence problems that are not easily resolved
(D.M. Stasinopoulos, personal communication).  However, the GAMLSS fit
is faster, and we have found the GAMLSS package to be very useful
for the exploratory examination of many potential models.

\subsection{Simulation studies}

We consider three simulation studies which show the
effectiveness of our methodology for detecting overdispersion when it exists
and for distinguishing between null, linear and flexible effects. We also
examine the gain in performance which results when our hierarchical variable
selection priors are used instead of a similar hierarchical prior in which
variable selection is not done. Performance here is measured by the
percentage increase in average Kullback-Leibler divergence between the true
and estimated predictive densities averaging over observed predictors.

For our fitted overdispersed model to the rats data, we
simulated 50 replicate datasets from the fitted model. Table \ref
{tab:ratsDispProbMeanVar} shows the $25$th and $75$th percentiles of the
probabilities of null, linear and flexible effects for the two predictors in
the mean and dispersion models over the 50 replications. The results are
consistent with our fit to the original data, with appreciable probabilities
of linear and flexible effects for sample size in the mean model and sample
size and hemoglobin level in the variance model and an appreciable
probability for no effect for sample size in the mean model. Table \ref
{tab:ratsDispProbMeanVardbp} is similar to Table \ref
{tab:ratsDispProbMeanVar} but for the data based priors method of Yau 
\textit{et al.} (2003), implemented using a similar approach to that
discussed in Section 4.2. The results for the data based priors approach are
similar to those for our hierarchical priors for the mean model.  
Posterior probabilities of null effects in the variance model differ
in the two implementations, but this may be due to the fact that
selection is done separately on different predictors in the variance
model in the data based priors approach, whereas we include or exclude
predictors together in our hierarchical prior with a shrinkage prior
on coefficients.  
Note that the results of Table \ref
{tab:ratsDispProbMeanVardbp} show only posterior probabilities for flexible
and null effects, as flexible and linear terms are selected together in the
approach of Yau \textit{et al.} (2003). Tables \ref
{tab:ratsDispProbMeanVarotherst1} and \ref{tab:ratsDispProbMeanVarotherst2}
show the results of the simulation study for our method where the
hyperparameters $s$ and $t$ in the inverse gamma priors of Section 2.2 are $%
(s,t)=(6,500)$ and $(s,t)=$ {\normalsize $(27,1300)$ (giving prior means of
100 and 50 respectively and standard deviations of 50 and 10 respectively).
As can be seen from the tables, the results of our approach are not
particularly sensitive to the choice of $s$ and $t$. 

\begin{table}[ht!]
\centering
\small{
\begin{tabular}{|cc|cccc|}
\hline
&  & \multicolumn{4}{c|}{Covariate} \\ 
&  & \multicolumn{2}{c}{Mean} & \multicolumn{2}{c|}{Var} \\ 
Effect & Percentile & Hem & SS & Hem & SS \\ \hline
Flexible & ${25}^{{th}}$ & 0.42 & 0.15 & 0.38 & 0.38 \\ 
& ${75}^{{th}}$ & 0.47 & 0.27 & 0.45 & 0.45 \\ \hline
Linear & ${25}^{{th}}$ & 0.52 & 0.19 & 0.55 & 0.55 \\ 
& ${75}^{{th}}$ & 0.58 & 0.30 & 0.61 & 0.61 \\ \hline
Null & ${25}^{{th}}$ & 0.00 & 0.44 & 0.00 & 0.00 \\ 
& ${75}^{{th}}$ & 0.00 & 0.64 & 0.00 & 0.00 \\ \hline
\end{tabular}
}
\vspace{10pt}
\caption{Rats data simulated from the fitted model. The $25$th and $75$th
percentiles of the probabilities of flexible, linear and null effects for
the mean and variance components, for the two covariates hemoglobin (Hem)
and sample size (SS). Based on 50 replications}
\label{tab:ratsDispProbMeanVar}
\end{table}
 
\begin{table}[ht!]
\centering 
\begin{tabular}{|cc|cccc|}
\hline
&  & \multicolumn{4}{c|}{Covariate} \\ 
&  & \multicolumn{2}{c}{Mean} & \multicolumn{2}{c|}{Var} \\ 
Effect & Percentile & Hem & SS & Hem & SS \\ \hline
Flexible & ${25}^{{th}}$ & 1.00 & 0.40 & 0.25 & 0.18 \\ 
& ${75}^{{th}} $ & 1.00 & 0.48 & 0.36 & 0.25 \\ \hline
Null & ${25}^{{th}} $ & 0.00 & 0.52 & 0.64 & 0.75 \\ 
& ${75}^{{th}} $ & 0.00 & 0.60 & 0.75 & 0.82 \\ \hline
\end{tabular}
\vspace{10pt}
\caption{Rats data simulated from the fitted model. The $25$th and $75$th
percentiles of the posterior probabilities of flexible and null effects for
the mean and variance components, for the two covariates hemoglobin (Hem)
and sample size (SS) for the data based priors approach. Based on 50
replications.}
\label{tab:ratsDispProbMeanVardbp}
\end{table}

\begin{table}[ht!]
\centering
\begin{tabular}{|cc|cccc|}
\hline
&  & \multicolumn{4}{c|}{Covariate} \\ 
&  & \multicolumn{2}{c}{Mean} & \multicolumn{2}{c|}{Var} \\ 
Effect & Percentile & Hem & SS & Hem & SS \\ \hline
Flexible & ${25}^{{th}}$ & $0.10$ & $0.04$ & $0.11$ & $0.10$ \\ 
& ${75}^{{th}} $ & $0.13$ & $0.06$ & $0.13$ & $0.13$ \\ \hline
Linear & ${25}^{{th}} $ & $0.86$ & $0.27$ & $0.87$ & $0.88$ \\ 
& ${75}^{{th}} $ & $0.89$ & $0.33$ & $0.89$ & $0.90$ \\ \hline
Null & ${25}^{{th}} $ & $0.00$ & $0.60$ & $0.00$ & $0.00$ \\ 
& ${75}^{{th}} $ & $0.01$ & $0.69$ & $0.00$ & $0.00$ \\ \hline
\end{tabular}
\vspace{10pt}
\caption{Rats data simulated from the fitted model. The $25$th and $75$th
percentiles of the probabilities of flexible, linear and null effects for
the mean and variance components, for the two covariates hemoglobin (Hem)
and sample size (SS). Based on 50 replications and hyperparameter settings $%
(s,t)=(27,1300)$.}
\label{tab:ratsDispProbMeanVarotherst1}
\end{table}

\begin{table}[ht!]
\centering
\begin{tabular}{|cc|cccc|}
\hline
&  & \multicolumn{4}{c|}{Covariate} \\ 
&  & \multicolumn{2}{c}{Mean} & \multicolumn{2}{c|}{Var} \\ 
Effect & Percentile & Hem & SS & Hem & SS \\ \hline
Flexible & ${25}^{{th}}$ & $0.09$ & $0.03$ & $0.11$ & $0.10$ \\ 
& ${75}^{{th}}$ & $0.12$ & $0.05$ & $0.14$ & $0.13$ \\ \hline
Linear & ${25}^{{th}}$ & $0.86$ & $0.23$ & $0.86$ & $0.87$ \\ 
& ${75}^{{th}}$ & $0.90$ & $0.28$ & $0.89$ & $0.90$ \\ \hline
Null & ${25}^{{th}}$ & $0.00$ & $0.66$ & $0.00$ & $0.00$ \\ 
& ${75}^{{th}}$ & $0.02$ & $0.73$ & $0.00$ & $0.00$ \\ \hline
\end{tabular}
\vspace{10pt}
\caption{Rats data simulated from the fitted model. The $25$th and $75$th
percentiles of the probabilities of flexible, linear and null effects for
the mean and variance components, for the two covariates hemoglobin (Hem)
and sample size (SS). Based on 50 replications and hyperparameter settings $%
(s,t)=(6,500)$.}
\label{tab:ratsDispProbMeanVarotherst2}
\end{table}

Table \ref{tab:ratsNoDispProbMeanVar} is similar to Table \ref
{tab:ratsDispProbMeanVar}, but for 50 replicate datasets simulated from a
fitted binomial model (that is, with no overdispersion). The probability of
a null effect for both covariates in the variance model is near one and
again there is high probability of a null effect for sample size in the mean
model and appreciable probabilities for linear and flexible effects for
hemoglobin in the mean model. Table \ref{tab:ratsNoDispProbMeanVardbp} is
similar to Table \ref{tab:ratsNoDispProbMeanVar} except for the data based
priors approach of Yau \textit{et al.} (2003). The results are again 
similar. 

\begin{table}[ht!]
\centering 
\begin{tabular}{|cc|cccc|} \hline
&  & \multicolumn{4}{c|}{Covariate} \\ 
&  & \multicolumn{2}{c}{Mean} & \multicolumn{2}{c|}{Var} \\ 
Effect & Percentile & Hem & SS & Hem & SS \\ \hline
Flexible & ${25}^{{th}}$ & 0.40 & 0.08 & 0.00 & 0.00 \\ 
& ${75}^{{th}}$ & 0.43 & 0.22 & 0.00 & 0.00 \\ \hline
Linear & ${25}^{{th}}$ & 0.57 & 0.12 & 0.00 & 0.00 \\ 
& ${75}^{{th}}$ & 0.60 & 0.36 & 0.00 & 0.00 \\ \hline
Null & ${25}^{{th}}$ & 0.00 & 0.43 & 0.99 & 0.99 \\ 
& ${75}^{{th}}$ & 0.00 & 0.79 & 0.99 & 0.99 \\ \hline
\end{tabular}
\vspace{10pt}
\caption{Rats simulated data from a fitted binomial model with no
overdispersion. The $25$th and $75$th percentiles of the probabilities of
flexible, linear and null effects are given for the mean and variance
components, for the two covariates hemoglobin (Hem) and sample size (SS).
Based on 50 replications.}
\label{tab:ratsNoDispProbMeanVar}
\end{table}

\begin{table}[h]
\centering 
\begin{tabular}{|cc|cccc|}
\hline
&  & \multicolumn{4}{c|}{Covariate} \\ 
&  & \multicolumn{2}{c}{Mean} & \multicolumn{2}{c|}{Var} \\ 
Effect & Percentile & Hem & SS & Hem & SS \\ \hline\hline
Flexible & ${25}^{{th}}$ & $1.00$ & $0.47$ & $0.00$ & $0.00$ \\ 
& ${75}^{{th}}$ & $1.00$ & $0.51$ & $0.00$ & $0.00$ \\ \hline
Null & ${25}^{{th}}$ & $0.00$ & $0.49$ & $1.00$ & $1.00$ \\ 
& ${75}^{{th}}$ & $0.00$ & $0.53$ & $1.00$ & $1.00$ \\ \hline
\end{tabular}
\vspace{10pt}
\caption{Rats simulated data from a fitted binomial model with no
overdispersion. The $25$th and $75$th percentiles of the posterior
probabilities of flexible and null effects are given for the mean and
variance components, for the two covariates hemoglobin (Hem) and sample size
(SS), and the data based priors approach. Based on 50 replications.}
\label{tab:ratsNoDispProbMeanVardbp}
\end{table}

Similarly to Table \ref{tab:ratsNoDispProbMeanVar} we examine in
Tables \ref{tab:ratsNoDispProbMeanVarotherst1} and \ref
{tab:ratsNoDispProbMeanVarotherst2} the performance of our approach but with
the hyperparameter settings $(s,t)=(6,500)$ and $(s,t)=(27,1300)$. Again
we see that the results are not sensitive to the hyperparameter settings.

\begin{table}[h]
\centering
\begin{tabular}{|cc|cccc|}
\hline
&  & \multicolumn{4}{c|}{Covariate} \\ 
&  & \multicolumn{2}{c}{Mean} & \multicolumn{2}{c|}{Var} \\ 
Effect & Percentile & Hem & SS & Hem & SS \\ \hline
Flexible & ${25}^{{th}}$ & $0.10$ & $0.02$ & $0.00$ & $0.00$ \\ 
& ${75}^{{th}}$ & $0.14$ & $0.05$ & $0.02$ & $0.01$ \\ \hline
Linear & ${25}^{{th}}$ & $0.86$ & $0.10$ & $0.00$ & $0.00$ \\ 
& ${75}^{{th}}$ & $0.90$ & $0.19$ & $0.12$ & $0.13$ \\ \hline
Null & ${25}^{{th}}$ & $0.00$ & $0.75$ & $0.86$ & $0.86$ \\ 
& ${75}^{{th}}$ & $0.00$ & $0.87$ & $1.00$ & $1.00$ \\ \hline
\end{tabular}
\vspace{10pt}
\caption{Rats simulated data from a fitted binomial model with no
overdispersion. The $25$th and $75$th percentiles of the probabilities of
flexible, linear and null effects are given for the mean and variance
components, for the two covariates hemoglobin (Hem) and sample size (SS).
Based on 50 replications and hyperparameter settings $(s,t)=(6,500)$.}
\label{tab:ratsNoDispProbMeanVarotherst1}
\end{table}

\begin{table}[ht!]
\centering
\begin{tabular}{|cc|cccc|}
\hline
&  & \multicolumn{4}{c|}{Covariate} \\ 
&  & \multicolumn{2}{c}{Mean} & \multicolumn{2}{c|}{Var} \\ 
Effect & Percentile & Hem & SS & Hem & SS \\ \hline
Flexible & ${25}^{{th}}$ & $0.10$ & $0.02$ & $0.00$ & $0.00$ \\ 
& ${75}^{{th}}$ & $0.14$ & $0.05$ & $0.03$ & $0.03$ \\ \hline
Linear & ${25}^{{th}}$ & $0.86$ & $0.14$ & $0.01$ & $0.01$ \\ 
& ${75}^{{th}}$ & $0.90$ & $0.25$ & $0.23$ & $0.24$ \\ \hline
Null & ${25}^{{th}}$ & $0.00$ & $0.67$ & $0.73$ & $0.73$ \\ 
& ${75}^{{th}}$ & $0.00$ & $0.83$ & $0.99$ & $0.99$ \\ \hline
\end{tabular}
\vspace{10pt}
\caption{Rats simulated data from a fitted binomial model with no
overdispersion. The $25$th and $75$th percentiles of the probabilities of
flexible, linear and null effects are given for the mean and variance
components, for the two covariates hemoglobin (Hem) and sample size (SS).
Based on 50 replications and hyperparameter settings $(s,t)=(27,1300)$.}
\label{tab:ratsNoDispProbMeanVarotherst2}
\end{table}

Table \ref{tab:DiabProbMeanVar} shows probabilities of null,
linear and flexible effects for the eight covariates in the diabetes example
for 50 simulated replicate datasets from an additive model fitted to the
real data. Again the results are consistent with out fit to the full model,
with high probability of a null effect for covariates 1, 3, 4 and 5 (the
number of pregnancies, diastolic blood pressure, skin triceps skin fold
thickness and 2-Hour serum insulin respectively), an appreciable probability
of a linear effect for covariate 2 (plasma glucose concentration) and high
probabilities of nonlinear effects for covariates 6, 7 and 8 (body mass
index, diabetes pedigree function and age). Table \ref
{tab:DiabProbMeanVardbp} is similar to \ref{tab:DiabProbMeanVar} but for the
data based priors approach of Yau \textit{et al.} (2003). Again the results
are similar to those obtained using our hierarchical priors.

\begin{table}[ht!]
\centering 
\begin{tabular}{|cc|cccccccc|}
\hline
&  & \multicolumn{8}{c|}{Covariate} \\ 
Effect & Perc. & 1 & 2 & 3 & 4 & 5 & 6 & 7 & 8 \\ \hline
Flexible & ${25}^{{th}}$ & 0.03 & 0.55 & 0.06 & 0.05 & 0.04 & 0.62 & 0.35 & 
0.99 \\ 
& ${75}^{{th}}$ & 0.06 & 0.61 & 0.11 & 0.11 & 0.12 & 0.97 & 0.54 & 1.00 \\ 
\hline
Linear & ${25}^{{th}}$ & 0.03 & 0.39 & 0.04 & 0.03 & 0.04 & 0.03 & 0.19 & 
0.00 \\ 
& ${75}^{{th}}$ & 0.05 & 0.45 & 0.08 & 0.09 & 0.08 & 0.38 & 0.40 & 0.01 \\ 
\hline
Null & ${25}^{{th}}$ & 0.88 & 0.00 & 0.80 & 0.79 & 0.82 & 0.00 & 0.04 & 0.00
\\ 
& ${75}^{{th}}$ & 0.94 & 0.00 & 0.90 & 0.91 & 0.92 & 0.00 & 0.46 & 0.00 \\ 
\hline
\end{tabular}
\vspace{10pt}
\caption{Simulated data from the fitted diabetes model. The 25th and 75th
percentiles of the posterior probabilities of flexible, linear and null
effects are given. Based on 50 replications.}
\label{tab:DiabProbMeanVar}
\end{table}

\begin{table}[ht!]
\centering
\begin{tabular}{|cc|cccccccc|}
\hline
&  & \multicolumn{8}{c|}{Covariate} \\ 
Effect & Perc. & 1 & 2 & 3 & 4 & 5 & 6 & 7 & 8 \\ \hline
Flexible & ${25}^{{th}}$ & $0.36$ & $1.00$ & $0.35$ & $0.34$ & $0.28$ & $1.00
$ & $0.26$ & $1.00$ \\ 
& ${75}^{{th}}$ & $0.52$ & $1.00$ & $0.49$ & $0.49$ & $0.48$ & $1.00$ & $1.00
$ & $1.00$ \\ \hline
Null & ${25}^{{th}}$ & $0.48$ & $0.00$ & $0.51$ & $0.51$ & $0.52$ & $0.00$ & 
$0.00$ & $0.00$ \\ 
& ${75}^{{th}}$ & $0.64$ & $0.00$ & $0.65$ & $0.66$ & $0.72$ & $0.00$ & $0.74
$ & $0.00$ \\ \hline
\end{tabular}
\vspace{10pt}
\caption{Simulated data from the fitted diabetes model. The 25th and 75th
percentiles of the posterior probabilities of flexible and null effects are
given for the data based priors approach. Based on 50 replications.}
\label{tab:DiabProbMeanVardbp}
\end{table}

Similar to Table \ref{tab:DiabProbMeanVar} we examine in Tables 
\ref{tab:DiabProbMeanVarotherst1} and \ref{tab:DiabProbMeanVarotherst2} the
performance of our approach but with the hyperparameter settings $%
(s,t)=(6,500)$ and $(s,t)=(27,1300)$. Again we see that the results are not
particularly sensitive to the hyperparameter settings.

\begin{table}[ht!]
\centering 
\begin{tabular}{|cc|cccccccc|}
\hline
&  & \multicolumn{8}{c|}{Covariate} \\ 
Effect & Perc. & 1 & 2 & 3 & 4 & 5 & 6 & 7 & 8 \\ \hline
Flexible & ${25}^{{th}}$ & $0.04$ & $0.51$ & $0.03$ & $0.06$ & $0.03$ & $0.99
$ & $0.42$ & $1.00$ \\ 
& ${75}^{{th}}$ & $0.21$ & $0.64$ & $0.20$ & $0.14$ & $0.14$ & $1.00$ & $0.62
$ & $1.00$ \\ \hline
Linear & ${25}^{{th}}$ & $0.04$ & $0.36$ & $0.03$ & $0.05$ & $0.04$ & $0.00$
& $0.25$ & $0.00$ \\ 
& ${75}^{{th}}$ & $0.10$ & $0.49$ & $0.11$ & $0.10$ & $0.10$ & $0.01$ & $0.41
$ & $0.00$ \\ \hline
Null & ${25}^{{th}}$ & $0.67$ & $0.00$ & $0.69$ & $0.77$ & $0.78$ & $0.00$ & 
$0.00$ & $0.00$ \\ 
& ${75}^{{th}}$ & $0.92$ & $0.00$ & $0.94$ & $0.89$ & $0.93$ & $0.00$ & $0.19
$ & $0.00$ \\ \hline
\end{tabular}
\vspace{10pt}
\caption{Simulated data from the fitted diabetes model. The 25th and 75th
percentiles of the posterior probabilities of flexible, linear and null
effects are given. Based on 50 replications and the hyperparmaeter settings $%
(s,t)=(6,500)$.}
\label{tab:DiabProbMeanVarotherst1}
\end{table}

\begin{table}[ht!]
\centering 
\begin{tabular}{|cc|cccccccc|}
\hline
&  & \multicolumn{8}{c|}{Covariate} \\ 
Effect & Perc. & 1 & 2 & 3 & 4 & 5 & 6 & 7 & 8 \\ \hline
Flexible & ${25}^{{th}}$ & $0.06$ & $0.54$ & $0.07$ & $0.07$ & $0.07$ & $0.99
$ & $0.48$ & $0.98$ \\ 
& ${75}^{{th}}$ & $0.18$ & $0.69$ & $0.14$ & $0.17$ & $0.14$ & $1.00$ & $0.67
$ & $1.00$ \\ \hline
Linear & ${25}^{{th}}$ & $0.04$ & $0.31$ & $0.06$ & $0.04$ & $0.06$ & $0.00$
& $0.17$ & $0.00$ \\ 
& ${75}^{{th}}$ & $0.09$ & $0.46$ & $0.11$ & $0.10$ & $0.09$ & $0.01$ & $0.41
$ & $0.00$ \\ \hline
Null & ${25}^{{th}}$ & $0.72$ & $0.00$ & $0.75$ & $0.72$ & $0.78$ & $0.00$ & 
$0.00$ & $0.00$ \\ 
& ${75}^{{th}}$ & $0.89$ & $0.00$ & $0.86$ & $0.88$ & $0.86$ & $0.00$ & $0.17
$ & $0.01$ \\ \hline
\end{tabular}
\vspace{10pt}
\caption{Simulated data from the fitted diabetes model. The 25th and 75th
percentiles of the posterior probabilities of flexible, linear and null
effects are given. Based on 50 replications and the hyperparameter settings $%
(s,t)=(27,1300)$.}
\label{tab:DiabProbMeanVarotherst2}
\end{table}

We now compare the performance of our hierarchical variable
selection priors with the same prior but where all terms are flexible (that
is, no variable selection is carried out). Our measure of performance is the
Kullback-Leibler divergence, averaged over the observed covariates.

In estimating the true response distribution $p_{0}(y|x)$ using
an estimate $\hat{p}(y|x)$ where $x$ denotes the covariates, the
Kullback-Leibler divergence is defined as 
\begin{equation*}
KL(\hat{p}(\cdot |x),p_{0}(\cdot |x))=\int p_{0}(y|x)\log \left[ \frac{\hat{p}%
(y|x)}{p_{0}(y|x)}\right] dy.
\end{equation*}
We define the average Kullback-Leibler divergence as 
\begin{equation*}
AKLD(\hat{p},p_{0})=\frac{1}{n}\sum_{i=1}^{n}KL(\hat{p}(\cdot
|x_{i}),p_{0}(\cdot |x_{i}))
\end{equation*}
where $x_{i}$, $i=1,...,n$ denotes the observed predictors. Writing $\hat{p}%
^{V}(y|x)$ for the estimated predictive density at $x$ for the variable
selection prior and $\hat{p}^{NV}(y|x)$ for the estimated predictive density
at $x$ for the prior without variable selection we define the average
percentage increase in Kullback-Leibler loss for variable selection compared
to no variable selection as 
\begin{equation*}
APKL=\frac{AKLD(\hat{p}^{NV},p_{0})-AKLD(\hat{p}^{V},p_{0})}{AKLD(\hat{p}%
^{V},p_{0})}.
\end{equation*}
When $APKL$ is positive, the prior that allows for variable selection
outperforms the prior that does not allow variable selection. Table \ref
{tab:klpercentage} shows the $10^{th},25^{th},50^{th},75^{th}$ and $90^{th}$
precentiles of the $APKL$ for the 50 replicate data sets generated in our
simulation study for the diabetes data, the rats data when no overdispersion
is present and the rats data when overdispersion is present. The table shows
that the median percentage increase in $APKL$ is positive for all three cases,
indicating an improvement for using our hierarchical variable selection
prior compared to not doing variable selection. Furthermore, for the rats
data with no overdispersion, even the 10th percentile exceeds 28\%. 

\begin{table}[ht!]
\centering 
\begin{tabular}{|l|l|l|l|l|l|}
\hline
{Dataset} & \multicolumn{5}{|l|}{Percentiles} \\ \hline
& {\ }${10}^{{th}}$ & {\ }${25}^{{th}}$ & {\ }${50}^{{th}}$ & ${75}^{{th\ }}$
& ${90}^{{th}}$ \\ \hline
{Diabetes } & {\ -14.49} & {\ -1.17} & {\ 16.74} & {\ 40.85 } & {\ 64.14} \\ 
\hline
{Rats with overdispersion} & {\ -13.92 } & {\ -5.18} & {\ 9.08} & {\ 30.48}
& {\ 62.20} \\ \hline
{Rats with no overdispersion} & {\ 28.67} & {\ 67.55} & {\ 155.52} & {\
293.43} & {\ 734.54} \\ \hline
\end{tabular}
\vspace{10pt}
\caption{$10^{th},25^{th},50^{th},75^{th}$ and $90^{th}$ precentiles in the
percentage increase in Kullback-leibler divergence when no variable
selection is carried out compared to when variable selection is carried out.
The results are based on 50 replications for the diabetes data, rats data
when no overdispersion is present and rats data when overdispersion is
present.}
\label{tab:klpercentage}
\end{table}

\section{ Conclusion}

The article develops a general Bayesian framework for variable
selection and model averaging in generalized linear models that allows for
over or under dispersion. The priors and sampling are innovative and the
flexibility of the approach is demonstrated using a number of examples,
ranging from fully parametric to fully nonparametric.

There are a number of natural extensions to the work described
here. Although we have implemented our approach to flexible regression for
the mean and variance using the double exponential family of distributions,
it is easy to implement a similar approach using other distributional
families for overdispersed count data such as the beta-binomial and negative
binomial.  We have demonstrated use of the beta-binomial in one of our
real data examples.  Flexible modelling of multivariate data could also be easily
accommodated in our framework by incorporation of other kinds of random
effects apart from those involved in our nonparametric functional forms.
These and other extensions are the subject of ongoing research. 

\section*{Acknowledgements}

This work was supported by an Australian Research
Council Grant.  We thank Dr Mikis Stasinopoulos for a quick and helpful
response to some questions about the GAMLSS package.


\section{Appendix}

This section gives details of the sampling scheme in Section 3.
Most of the steps involve an application of the Metropolis-Hastings method
based on one of the following two approximations to the conditional
densities. 

\noindent 
\underline{Approximation 1 }

Here we seek to generate a parameter $\psi $ from its full
conditional $p(\psi |y,\Delta \backslash \psi ),$ where $\Delta $ consists
of all the parameters and latent variables used in the sampling scheme and $%
\Delta \backslash \psi $ means all of $\Delta $ excluding $\psi .$ We write 
\begin{eqnarray}
p(\psi |y,\Delta \backslash \psi ) &\propto &p(y|\psi ,\Delta \backslash
\psi )p(\psi |\Delta \backslash \psi )  \label{equ:postcondphi} \\
&\propto &\exp (-l(\psi ))  \notag
\end{eqnarray}
where $l(\psi )$ is the negative of the logarithm of the left side of (\ref
{equ:postcondphi}) and for convenience the dependence of $l(\cdot )$ on $%
\Delta \backslash \psi $ is omitted. 

Let $\hat{\psi}$ be the minimum of $l(\psi )$ and $\Psi
=\partial ^{2}l(\hat{\psi})/\partial \psi \partial \psi $. We approximate
the full conditional of $\psi $ by a normal density with mean $\hat{\psi}$
and covariance matrix $\Psi ^{-1}.$ Generally we find $\hat{\psi}$ using
numerical optimization routines from the NAG or IMSL libraries and have not
experienced any difficulties of convergence. As a practical matter, there
may be a substantial benefit to early stopping of the optimization in
constructing our proposals after just a few or even one step. Just a few
steps gives a good approximation to $\hat{\psi}$ and this suffices to obtain
good proposals with the early stopping resulting in a considerable saving in
computation time since several applications of Approximation 1 are done at
every iteration of the sampling scheme. 

\noindent 
\underline{Approximation 2 }

Here we seek to generate parameters $\psi $ and $w$ as a block
from their joint conditional density. We assume that $p(y|\psi ,w,\Delta
\backslash \left\{ \psi ,w\right\} )=p(y|\psi ,\Delta \backslash \left\{
\psi ,w\right\} )$ and $p(\psi |w,\Delta \backslash \left\{ \psi ,w\right\}
)=$ $p(\psi |w)$ is Gaussian. Then, 
\begin{eqnarray*}
p(w|y,\Delta \backslash \left\{ \psi ,w\right\} ) &=&\int p(\psi ,w|y,\Delta
\backslash \left\{ \psi ,w\right\} )d\psi \\
&\propto &\int p(y|\psi ,\Delta \backslash \left\{ \psi ,w\right\} )p(\psi
|w)d\psi \times p(w|\Delta \backslash \left\{ \psi ,w\right\} ).
\end{eqnarray*}
Let $q_{1}(\psi )$ be a Gaussian approximation to $p(y|\psi ,\Delta
\backslash \left\{ \psi ,w\right\} )$ as in approximation 1, and note that
it is independent of $w.$ Let 
\begin{equation*}
q_{2}(w)=\int q_{1}(\psi )p(\psi |w)d\psi
\end{equation*}
and note that $q_{2}(w)$ can be evaluated explicitly as a function of $w.$
Precisely, if $q_1(\psi)$ is a Gaussian $N(\mu_1,\Sigma_1)$ and $p(\psi|w)$
is a Gaussian $N(\mu_2,\Sigma_w)$ then we obtain 
\begin{eqnarray*}
q_2(w) & \propto &
|\Sigma_w^{-1}|^{-1/2}|\Sigma_1^{-1}+\Sigma_w^{-1}|^{-1/2} \exp\left(-\frac{1%
}{2}\left\{\mu_w^T\Sigma_w^{-1}\mu_w+\right.\right. \\
& &
\left.\left.(\mu_1^T\Sigma_1^{-1}+\mu_w^T\Sigma_w^{-1})^T(\Sigma_1^{-1}+%
\Sigma_w^{-1})^{-1}
(\mu_1^T\Sigma_1^{-1}+\mu_w^T\Sigma_w^{-1})\right\}\right).
\end{eqnarray*}
Thus, we approximate $p(w|y,\Delta \backslash \left\{ \psi ,w\right\} )$ by $%
q_{3}(w)=q_{2}(w)p(w|\Delta \backslash \left\{ \psi ,w\right\} ).$ In our
applications $w$ is scalar and so it is straightforward to approximate $%
q_{3}(w)$ by a Gaussian as in the first approximation and hence generate $w$%
. Once $w$ is generated, it is straightforward to generate $\psi $ from $%
p(\psi |w)$ which is Gaussian. 

In step 1, let $\psi =\beta _{0}^{\mu }$ and construct a
proposal density for $\psi $ as in the first approximation. We either accept
the proposed value or retain the current value according to the usual
Metropolis-Hastings rule. 

In Step 2, $J_{j}^{\mu }$ and $\beta _{j}^{\mu }$ are generated
as a block. If $K_{j}^{\mu }=1$ then $J_{j}^{\mu }=1.$ If $K_{j}^{\mu }=0$
then $J_{j}^{\mu }$ is generated as $0$ or $1$ from the prior. If $%
J_{j}^{\mu }$ is generated as a $1,$ then $\beta _{j}^{\mu }$ is generated
from a normal approximation to its full conditional density as in the first
approximation. The proposed pair is either accepted or it is rejected in
favor of the current values according to the Metropolis-Hastings rule.

In step 3, $b^{\mu }$ is sampled from its full conditional
density, 
\begin{equation}
p(b^{\mu }|y,\Delta \backslash b^{\mu })\propto p(\beta ^{\mu }|b^{\mu
})p(b^{\mu })  \label{e:bmufullcond}
\end{equation}
the right side of (\ref{e:bmufullcond}) is the unnormalized inverse gamma
density with shape parameter $s+\sum_{j}J_{j}^{\mu }/2$ and scale parameter $%
t+{\beta _{J}^{\mu }}^{T}\beta _{J}^{\mu }/2.$

To describe step 4 of the sampling scheme, we first show how $%
c_{j}^{\mu }$ and $\alpha _{j}^{\mu }$ are generated if $K_{j}^{\mu }=1.$
Let $\psi =\alpha _{j}^{\mu }$ and $w=c_{j}^{\mu }.$ Then $p(y|\psi
,w,\Delta \backslash \left\{ \psi ,w\right\} )=p(y|\psi ,\Delta \backslash
\left\{ \psi ,w\right\} )$ and $p(\psi |w)$ is Gaussian in $\psi .$ We
generate $(\psi ,w)$ as a block as in the second approximation.

We generate a proposal for $(K_{j}^{\mu },c_{j}^{\mu },\alpha
_{j}^{\mu })$ as follows. First, generate a proposal for $K_{j}^{\mu }$ as $%
0 $ if $J_{j}^{\mu }=0,$ and $0$ or $1$ from the prior if $J_{j}^{\mu }=1.$
Next, if the proposed value of $K_{j}^{\mu }=1$ then generate $(c_{j}^{\mu
},\alpha _{j}^{\mu })$ as outlined above. Then, we accept or reject the
block proposal according to the Metropolis-Hastings rule.

In step 5, $a^{c\mu }$ and $b^{c\mu }$ can be updated from
their full conditional distributions. We have 
\begin{equation*}
p(a^{c\mu }|\Delta \backslash a^{c\mu })\propto p(a^{c\mu })p(c^{\mu
}|a^{c\mu },b^{c\mu })
\end{equation*}
and we recognize the right hand side as an unnormalized normal density. We
have 
\begin{equation*}
p(a^{c\mu }|\Delta \backslash \{a^{c\mu }\})=N\left( \left( \frac{%
\sum_{j}K_{j}^{\mu }}{b^{c\mu }}+\frac{1}{100}\right) ^{-1}\frac{(c^{\mu
})^{T}c^{\mu }}{b^{c\mu }},\left( \frac{\sum_{j}K_{j}^{\mu }}{b^{c\mu }}+%
\frac{1}{100}\right) ^{-1}\right) .
\end{equation*}
Also, 
\begin{equation*}
p(b^{c\mu }|\Delta \backslash \{b^{c\mu }\})\propto p(b^{c\mu })p(c^{\mu
}|a^{c\mu },b^{c\mu })
\end{equation*}
and we recognize the right hand side as an unnormalized inverse gamma
density. We have 
\begin{equation*}
p(b^{c\mu }|\Delta \backslash \{b^{c\mu }\})=IG\left( s+\frac{%
\sum_{j}K_{j}^{\mu }}{2},t+\frac{1}{2}(c^{\mu }-a^{c\mu })^{T}(c^{\mu
}-a^{c\mu })\right) .
\end{equation*}
Step 6 is similar to step 1, except that a multivariate normal approximation
is used to generate $\beta ^{\theta }$ given $J^{\theta }=1$ and the current
values of other parameters. Step 7 is similar to step 3 and 
\begin{equation*}
p(b^{\theta }|\Delta \backslash \{b^{\theta },J^{\theta }\},J^{\theta
}=1)=IG\left( s+\frac{q+1}{2},t+\frac{1}{2}\left( {\beta ^{\theta }}\right)
^{T}\beta ^{\theta }\right) .
\end{equation*}
Steps 8 and 9 are performed similarly to steps 4 and 5.


\begin{thebibliography}{37}
\newcommand{\enquote}[1]{``#1''}
\expandafter\ifx\csname natexlab\endcsname\relax\def\natexlab#1{#1}\fi

\bibitem[{Aerts and Claskens(1997)}]{aerts97}
Aerts, M. and Claskens, G. (1997), \enquote{Local polynomial estimation in
  multiparameter models,} \textit{Journal of the American Statistical
  Association}, 92, 1536--1545.

\bibitem[{Breslow(1990)}]{breslow90}
Breslow, N. (1990), \enquote{Further studies in the variability of pock
  counts,} \textit{Statistics in Medicine}, 9, 615--626.

\bibitem[{Breslow and Clayton(1993)}]{breslow93}
Breslow, N. and Clayton, D. (1993), \enquote{Approximate inference in
  generalized linear mixed models,} \textit{Journal of the American Statistical
  Association}, 88, 9--25.

\bibitem[{Brezger and Lang(2005)}]{brezger05}
Brezger, A. and Lang, S. (2005), \enquote{Generalized additive structured
  regression based on Bayesian P-splines,} \textit{Computational Statistics and
  Data Analysis}, 50, 967--991.

\bibitem[{Davidian and Carroll(1987)}]{davidian87}
Davidian, M. and Carroll, R. (1987), \enquote{Variance function estimation,}
  \textit{Journal of the American Statistical Association}, 82, 1079--1091.

\bibitem[{Davidian and Carroll(1988)}]{davidian88}
--- (1988), \enquote{A note on extended quasilikelihood,} \textit{Journal of
  the Royal Statistical Society {B}}, 50, 74--82.

\bibitem[{Davidian and Giltinan(1995)}]{davidian95}
Davidian, M. and Giltinan, D. (1995), \textit{Nonlinear Models for Repeated
  Measurement Data}, New York: Chapman and Hall.

\bibitem[{Efron(1986)}]{efron86}
Efron, B. (1986), \enquote{Double exponential families and their use in
  generalised linear regression,} \textit{Journal of the American Statistical
  Association}, 81, 709--721.

\bibitem[{Eilers and Marx(1996)}]{eilers96}
Eilers, P. H.~C. and Marx, B.~D. (1996), \enquote{Flexible Smoothing with
  {B}-splines and Penalties with Rejoinder,} \textit{Statistical Science}, 11,
  89--121.

\bibitem[{Faddy(1997)}]{faddy97}
Faddy, M. (1997), \enquote{Extended Poisson process modelling and analysis of
  count data,} \textit{Biometrical Journal}, 39, 431--440.

\bibitem[{Gelfand and Dalal(1990)}]{Gelfand90a}
Gelfand, A. and Dalal, S. (1990), \enquote{A note on overdispersed exponential
  families,} \textit{Biometrika}, 77, 55--64.

\bibitem[{Gelfand et~al.(1997)Gelfand, Dey, and Peng}]{Gelfand97}
Gelfand, A., Dey, D., and Peng, F. (1997), \enquote{Overdispersed generalized
  linear models,} \textit{Journal of Statistical Planning and Inference}, 64,
  93--107.

\bibitem[{Gu(2002)}]{gu02}
Gu, C. (2002), \textit{Smoothing spline {ANOVA} models}, New York:
  Springer-Verlag.

\bibitem[{Hastie(1996)}]{hastie96}
Hastie, T. (1996), \enquote{Pseudosplines,} \textit{Journal of the Royal
  Statistical Society {B}}, 58, 379--396.

\bibitem[{Hastie and Tibshirani(1990)}]{hastie90}
Hastie, T. and Tibshirani, R. (1990), \textit{Generalized Additive Models}, New
  York: Chapman and Hall.

\bibitem[{Jorgensen(1997)}]{jorgensen97}
Jorgensen, B. (1997), \textit{The Theory of Dispersion Models}, London: Chapman
  and Hall.

\bibitem[{Lee and Nelder(1996)}]{lee96}
Lee, Y. and Nelder, J. (1996), \enquote{Hierarchical generalized linear models
  (with {D}iscussion),} \textit{Journal of the Royal Statistical Society {B}},
  58, 619--678.

\bibitem[{Lin and Zhang(1999)}]{lin99}
Lin, X. and Zhang, D. (1999), \enquote{Inference in generalized additive mixed
  models by using smoothing splines,} \textit{Journal of the Royal Statistical
  Society {B}}, 61, 381--400.

\bibitem[{Liu(2001)}]{liu01}
Liu, J. (2001), \textit{Monte Carlo Strategies in Scientific Computing}, New
  York: Springer-Verlag.

\bibitem[{McCullagh and Nelder(1989)}]{MCcullagh89}
McCullagh, P. and Nelder, J. (1989), \textit{Generalized Linear Models},
  London: Chapman and Hall, 2nd ed.

\bibitem[{Moore and Tsiatis(1991)}]{moore91}
Moore, D. and Tsiatis, A. (1991), \enquote{Robust estimation of the variance in
  moment methods for extra-binomial and extra-Poisson variation,}
  \textit{Biometrics}, 47, 383--401.

\bibitem[{Nelder and Pregibon(1987)}]{nelder87}
Nelder, J. and Pregibon, D. (1987), \enquote{An extended quasi-likelihood
  function,} \textit{Biometrika}, 74, 221--232.

\bibitem[{Nelder and Wedderburn(1972)}]{nelder72}
Nelder, J. and Wedderburn, R. (1972), \enquote{Generalized linear models,}
  \textit{Journal of the Royal Statistical Society {B}}, 135, 370--384.

\bibitem[{Nott(2004)}]{nott04}
Nott, D. (2004), \enquote{Semiparametric estimation of mean and variance
  functions for non-{G}aussian data,} Working paper.

\bibitem[{Peck et~al.(1984)Peck, Beal, Sheiner, and Nichols}]{peck84}
Peck, C., Beal, S., Sheiner, L., and Nichols, A. (1984), \enquote{Extended
  least squares nonlinear regression: A possible solution to the choice of
  weights problem in analysis of individual pharmacokinetic data,}
  \textit{Journal of Pharmacokinetics and Biopharmaceutics}, 12, 545--558.

\bibitem[{Podlich et~al.(2004)Podlich, Faddy, and Smyth}]{Podlich04}
Podlich, H., Faddy, M., and Smyth, G. (2004), \enquote{Semi-parametric extended
  {P}oisson process models,} \textit{Statistics and Computing}, 14, 311--321.

\bibitem[{Rigby and Stasinopoulos(2005)}]{rigby05}
Rigby, R. and Stasinopoulos, D. (2005), \enquote{Generalized additive models
  for location, scale and shape,} \textit{Applied Statistics}, 54, 1--38.

\bibitem[{Ruppert et~al.(2003)Ruppert, Wand, and Carroll}]{Ruppert03}
Ruppert, D., Wand, M., and Carroll, R. (2003), \textit{Semiparametric
  regression}, Cambridge: Cambridge University Press.

\bibitem[{Shively et~al.(1999)Shively, Kohn, and Wood}]{shively99}
Shively, S., Kohn, R., and Wood, S. (1999), \enquote{Variable selection and
  function estimation in additive nonparametric regression using a data based
  prior (with discussion),} \textit{Journal of the American Statistical
  Association}, 94, 777--807.

\bibitem[{Smith and Kohn(1996)}]{Smith96}
Smith, M. and Kohn, R. (1996), \enquote{Nonparametric Regression Using
  {B}ayesian Variable Selection,} \textit{Journal of Econometrics}, 75,
  317--344.

\bibitem[{Smyth(1989)}]{smyth89}
Smyth, G. (1989), \enquote{Generalized linear models with varying dispersion,}
  \textit{Journal of the Royal Statistical Society {B}}, 51, 47--60.

\bibitem[{Smyth and Verbyla(1999)}]{smyth99}
Smyth, G.~K. and Verbyla, A.~P. (1999), \enquote{Adjusted likelihood methods
  for modelling dispersion in generalized linear models,}
  \textit{Environmetrics}, 10, 696--709.

\bibitem[{Wahba et~al.(1995)Wahba, Wang, Gu, Klein, and Klein}]{wahba95}
Wahba, G., Wang, Y., Gu, C., Klein, R., and Klein, B. (1995),
  \enquote{Smoothing spline {ANOVA} for exponential families, with application
  to the Wisconsin Epidemiological Study of Diabetic Retinopathy,}
  \textit{Annals of Statistics}, 23, 1865--1895.

\bibitem[{Wedderburn(1974)}]{wedderburn74}
Wedderburn, R. (1974), \enquote{Quasi-likelihoods functions, generalized linear
  models and the Gauss-Newton method,} \textit{Biometrika}, 61, 439--447.

\bibitem[{Wild and Yee(1996)}]{wild96}
Wild, C. and Yee, T. (1996), \enquote{Additive extensions to generalized
  estimating equation methods,} \textit{Journal of the Royal Statistical
  Society {B}}, 58, 711--725.

\bibitem[{Yau et~al.(2003)Yau, Kohn, and Wood}]{yau03a}
Yau, P., Kohn, R., and Wood, S. (2003), \enquote{Bayesian variable selection
  and model averaging in high-dimensional multinomial nonparametric
  regression,} \textit{Journal of Computational and Graphical Statistics}, 12,
  23--54.

\bibitem[{Yee and Wild(1996)}]{Yee96}
Yee, T. and Wild, C. (1996), \enquote{Vector Generalized Additive Models,}
  \textit{Journal of the Royal Statistical Society {B}}, 58, 481--493.

\end{thebibliography}
\end{document}